\documentclass[10pt,onecolumn,preprintnumbers]{revtex4}
\pdfoutput=1
\usepackage{graphicx}

\usepackage[T1]{fontenc}
\usepackage{ae,aecompl}
\usepackage{amsmath,amsfonts,amssymb,mathrsfs}
\usepackage{tikz}
\usetikzlibrary{calc}
\usepackage{pgfplots}

\usepackage{xspace}

\usepackage{hyperref}
\hypersetup{colorlinks=true,linkcolor=blue,pdfstartview=XYZ}
\hypersetup{
colorlinks=false,
citecolor=green,
linkcolor=blue,
urlcolor=purple
}

\usepackage[titletoc]{appendix}
\usepackage{cleveref}

\usepackage{subfigure}

\def\laq{~\raise 0.4ex\hbox{$<$}\kern -0.8em\lower 0.62ex\hbox{$\sim$}~}
\def\gaq{~\raise 0.4ex\hbox{$>$}\kern -0.7em\lower 0.62ex\hbox{$\sim$}~}

\def\beq{\begin{equation}}
\def\eeq{\end{equation}}
\def\bea{\begin{eqnarray}}
\def\eea{\end{eqnarray}}

\def \ti {\widetilde}

\newcommand{\Ups}{\Upsilon}

\newcommand{\tbf}[1]{\textbf{#1}}

\newcommand{\rref}[1]{(\ref{#1})}
\newcommand{\di}{\mathrm{d}}

\newcommand{\Scal}{\mathcal S}
\newcommand{\Acal}{\mathcal A}

\newcommand{\tdev}{\mbox{\Large $\cdot$}}

\begin{document}

\preprint{BA-TH/682-13}

\title{Lensing in the Geodesic Light-Cone coordinates \\ and its (exact) illustration to an off-center observer in Lema\^itre-Tolman-Bondi models}

\author{G. Fanizza$^{1,2}$\footnote{giuseppe.fanizza@ba.infn.it}, F. Nugier$^{3,4}$\footnote{fabienjean.nugier@unibo.it}}

\affiliation{$^{1}$Dipartimento di Fisica, Universit\`{a} di Bari, Via G. Amendola
173, 70126 Bari, Italy\\
$^{2}$Istituto Nazionale di Fisica Nucleare, Sezione di Bari, Bari, Italy\\
$^{3}$Glenco Group, Dipartimento di Fisica e Astronomia, Universit\`{a} di Bologna, viale B. Pichat 6/2 , 40127, Bologna, Italy\\
$^{4}$Istituto Nazionale di Fisica Nucleare, Sezione di Bologna, Bologna, Italy\\
}

\begin{abstract}
We present in this paper a new application of the geodesic light-cone (GLC) gauge for weak lensing calculations. Using interesting properties of this gauge, we derive an exact expression of the amplification matrix -- involving convergence, magnification and shear -- and of the deformation matrix -- involving the optical scalars. These expressions are simple and non-perturbative as long as no caustics are created on the past light-cone and are, by construction, free from the thin lens approximation. We apply these general expressions on the example of an Lema\^itre-Tolman-Bondi (LTB) model with an off-center observer and obtain explicit forms for the lensing quantities as a direct consequence of the non-perturbative transformation between GLC and LTB coordinates. We show their evolution in redshift after a numerical integration, for underdense and overdense LTB models, and interpret their respective variations in the simple non-curvature case.

\end{abstract}

\vspace {1cm}~


\maketitle

\setcounter{equation}{0}

\section{Introduction}
\label{Sec1}
\label{SecIntro}

Lensing plays a very significant role in today's cosmological experiments, e.g. \cite{Ade:2013tyw, Hoekstra:2013gua,2012ApJ...756..142Vcut,Das:2013zf}, and will certainly receive a growing interest in the next decades \cite{Amendola:2012ys, Abbott:2005bi, 2009arXiv0912.0201L}. One explanatory reason for that is that lensing is a phenomenon happening on many scales in the Universe, from (strong) microlensing acting at micro-arcsecond angles where we use stellar objects as lenses, or giant arcs of the order of the arcsecond around clusters of galaxies, to the order of several arcminutes in the CMB. On the other hand the theory of lensing is known since a long time \cite{hawking1973large,1986ApJ...310..568B, schneider1999gravitational, petters2001singularity} and a lot of developments have been given to that field of research \cite{Uzan:2000xv, Lewis:2006fu, Fleury:2014gha}.

Nevertheless, descriptions of lensing usually make assumptions \cite{petters2001singularity}. One well known assumption is the thin lens approximation in which the lenses are assumed to be of a negligible size compared to the length of photon paths. The use of the Jacobi map formalism, as employed in this paper, is independent from this assumption and hence can be seen as more general. It depends on the other hand on the Born and geometrical optics approximations, i.e. that the angles of deviation are small (in practice typically less than arcminutes) and the wavelength of light is irrelevant. Similarly, one often assumes that there are no caustics on the sky (in which case the magnification is infinite and previous assumptions can break down), namely that the lenses under study are not `strong'. This is a stronger assumption than the small-angles/Born approximation as small size sources can develop caustics even with small angles.

In a recent set of papers, \cite{P1,P2,P3,P4,P5,Fanizza:2013doa,Marozzi:2014kua,DiDio:2014lka}, a system of coordinates -- the \emph{geodesic light-cone (GLC) coordinates} -- has been employed to derive expressions of observational quantities in a non-perturbative and possibly new interpretational way. These coordinates, which are adapted to the propagation of photons along the past light-cone of an observer, have proved themselves useful for the computation of the redshift, the luminosity distance \cite{P1} and the Jacobi map \cite{Fanizza:2013doa}.
We present in this paper a new application of this system of coordinates for the case of lensing. The expressions of the amplification and deformation matrices are derived and rely on the Jacobi map expressed in terms of zweibeins parallely transported along null geodesics. These matrices hence rely on the geometrical optics approximation \cite{petters2001singularity} and, as mentioned above, are general as long as no caustic forms on the past light-cone. We derive the lensing quantities they contain within the GLC coordinates, the main results of this paper, and show that they take relatively simple forms. We then discuss the specific example of a Lema\^itre-Tolman-Bondi (LTB) geometry present on our past light-cone, as an illustration of our previous results. The corresponding lensing quantities are derived exactly and their evolution in terms of redshift is shown after a numerical integration in the LTB coordinates for over/underdensities present in the Universe, illustrating the broad range of application of our expressions.

The paper is organised as follows. In Sec. \ref{Sec2} we give general expressions concerning mostly the propagation of the Jacobi map along null geodesics. We also introduce the amplification and deformation matrices with the lensing quantities they contain. In Sec. \ref{Sec3} we make use of the GLC coordinates to express the Jacobi map in a simple form. This expression involves zweibeins of the Sachs basis transported along the null geodesics and we show that some combinations of the lensing quantities do not necessitate their explicit form. We are hence able to derive non-perturbative expressions for these latter quantities within the GLC gauge. In Sec. \ref{Sec4} we present a general LTB model, off-centered with respect to the past light-cone of the considered observer, and find the non-perturbative transformation between GLC and LTB coordinates. This authorises us to derive in the rest of this section the exact expressions for lensing quantities as seen by an off-center observer in an LTB model. This is, for this geometry and up to our knowledge, the most explicit derivation of lensing quantities presented in the literature. Finally, in Sec. \ref{Sec5} we address the particular case of an LTB over/underdensity present in our past light-cone, giving more explicit formulas (but inevitably depending on a numerical integration when expressing them in terms of redshift) and physical interpretations for cosmology. We conclude our paper in Sec. \ref{Sec6} and present some Appendices (\ref{SecAppendixA}, \ref{SecAppendixB} and \ref{SecAppendixC}) relevant for our considerations (including the explicit form of the zweibeins in the GLC coordinates).

\section{Jacobi map, Amplification matrix and Deformation matrix - General definitions}
\label{Sec2}

Let us start by considering two light rays which are emitted at the same time from a source $S$ and which converge to an observer $O$. At each time, their relative separation is described by the geodesic deviation equation\,:
\beq
\nabla_\lambda^2 \xi^\mu = R_{\alpha\beta\nu}^{\mu} k^\alpha k^\nu \xi^\beta
\eeq
with $k^\mu$ the photon 4-momentum, $\nabla_\lambda\equiv {\rm D}/\di \lambda \equiv k^\mu \nabla_{\mu}$ with $\lambda$ an affine parameter along the photon path, and $\xi^\mu$ an orthogonal displacement with respect to the line of sight. As well known \cite{Pitrou:2012ge,Fleury:2013sna,Fanizza:2013doa}, this equation can be projected on a particular two dimensional spatial hypersurface thanks to the so-called Sachs basis $\{ s^\mu_A \}_{A=1,2}$ satisfying\,:
\bea
\label{eq:Sachs}
& g_{\mu\nu} s^\mu_A s^\nu_B = \delta_{AB} ~~,~~ s^\mu_A u_\mu = 0 ~~,~~ s^\mu_A k_\mu = 0 ~~, \nonumber\\
& \Pi^\mu_\nu\nabla_\lambda s^\nu_A = 0 ~~~~\text{with}~~~~\Pi^\mu_\nu=\delta^\mu_\nu-\frac{k^\mu k_\nu}{\left( u^\alpha k_\alpha \right)^2}-\frac{k^\mu u_\nu+u^\mu k_\nu}{u^\alpha k_\alpha} ~~,
\eea
where $u_\mu$ is the peculiar velocity of the comoving fluid and $\Pi^\mu_\nu$ is a projector on the two-dimensional space orthogonal to $u_\mu$ (the \emph{screen}) and $n_\mu= u_\mu+\left( u^\alpha k_\alpha \right)^{-1} k_{\mu}$ (with $n^\alpha n_\alpha =1$ and $n^\alpha k_\alpha=0$). Therefore, the projected quantities $\xi^A=\xi^\mu s^A_\mu$ and $R^A_B=R_{\alpha\beta\nu\mu}k^\alpha k^\nu s^\beta_B s^\mu_A$ allow us to obtain the (linear) $2^{\rm nd}$ order differential \emph{Jacobi equation}\,:
\beq
\label{EqEvolJAB}
\frac{\di^2}{\di \lambda^2} J^A_B (\lambda,\lambda_o) = R^A_C (\lambda) \, J^C_B (\lambda,\lambda_o) ~~,
\eeq
with initial conditions
\beq
\label{eq:initialConditions}
J^A_B(\lambda_o,\lambda_o) = 0 ~~~ \mbox{ and } ~~~ \frac{\di}{\di \lambda} J^A_B (\lambda_o,\lambda_o) = (k^\mu u_\mu)_o \, \delta^A_B ~~,
\eeq
once we have identified\,:
\beq
\label{DefinitionXisource}
\xi^A(\lambda) = J^A_B(\lambda,\lambda_o) \left( \frac{k^\mu \partial_\mu \xi^B}{k^\nu u_\nu} \right)_{\! \! o} ~~.
\eeq
The matrix $J^A_B$ is the so-called \emph{Jacobi map} and its link with observable quantities is well studied in literature \cite{hawking1973large, schneider1999gravitational, petters2001singularity} (more recently \cite{Fanizza:2013doa,Clarkson:2011br,Reimberg:2012uc,Yamauchi:2013fra}). In fact, the determinant of this map gives the angular distance of the considered source\,:
\beq
\label{dAJAB}
d_A(\lambda_s) = \sqrt{\det J^A_B(\lambda_s,\lambda_o)} ~~.
\eeq

The study of the Jacobi map is also relevant to understand the weak lensing effects on the image of an emitting source. To see this we have to define the relation between the unlensed angular position of the source $\bar\theta^A_s$ and the observed lensed position $\bar\theta^A_o$ (of the image). We choose these angles to be (similarly as \cite{schneider1999gravitational, Lewis:2006fu, Reimberg:2012uc} but unlike \cite{Uzan:2000xv})\,:
\bea
\label{DefinitionAngles}
\bar\theta^A_o=\left( \frac{k^\mu\partial_\mu\xi^A}{k^\mu u_\mu} \right)_o ~~~,~~~ \bar\theta^A_s=\left(\frac{\xi^A}{\bar d_A}\right)_s ~~.
\eea
The definition of the source's angular position is normalized with respect to the angular distance $\bar d_A$ of the homogeneous and isotropic background that our model is referring to. For instance, if we want to study the weak lensing due to inhomogeneities in a spatially non-flat background, we have to consider a spatially non-flat FLRW case for $\bar d_A$. This point will be clarified by the use of the GLC coordinates in Sec. \ref{Sec3} and in the illustration of the explicit LTB model of Sec. \ref{Sec4}.

From the definition of angles, Eq. \rref{DefinitionAngles}, the \emph{amplification matrix} (or \emph{lens mapping matrix}) is defined as follows\,:
\beq
\label{DefinitionA}
\Acal^A_B \equiv \frac{\di \bar{\theta}^A_s}{\di \bar{\theta}^B_o} = \frac{J^A_B (\lambda_s,\lambda_o)}{\bar d _A(\lambda_s)} ~~,
\eeq
where the last equality comes out by noticing that $\xi^A = J^A_B \, \bar\theta^B_o$ (see Eq. \rref{DefinitionXisource}). The physical meaning of this matrix is given by its decomposition in terms of a trace and a traceless part\,:
\beq
\label{MatrixA}
\Acal = \left( \begin{array}{cc} 1 - \kappa - \hat{\gamma}_1 & - \hat{\gamma}_2 + \hat{\omega} \\ - \hat{\gamma}_2 - \hat{\omega} & 1 - \kappa + \hat{\gamma}_1 \end{array} \right)\\
\eeq
where $\kappa$ is the \emph{dimensionless surface mass density} ($1 - \kappa$ is the \emph{convergence} but we will ``abusively'' use this latter term for $\kappa$), $|\hat\gamma|^2=\hat\gamma_1^2+\hat\gamma_2^2$ is the \emph{shear} and $\hat \omega$ the \emph{vorticity}. Furthermore, the determinant is an important quantity due to the definition of the \emph{magnification} $\mu=(\det\Acal)^{-1}$. In terms of the Jacobi map, all these objects can be expressed as\,:
\bea
\label{ExpressionsLensingQuantities}
& & \kappa = 1-\frac{\text{tr}J^A_B}{2\bar d_A}~~,~~\mu=\frac{\bar d_A^2}{\det J^A_B}~~,~~\hat\omega=\frac{|J^1_2-J^2_1|}{2\bar d_A} ~~, \nonumber \\
& & |\hat\gamma|^2 = \left(\frac{\text{tr}J^A_B}{2\bar d_A}\right)^2 + \left(\frac{|J^1_2-J^2_1|}{2\bar d_A} \right)^2 - \frac{\det J^A_B}{\bar d_A^2} ~~,
\eea
where for this last quantity we used $|\hat{\gamma} |^2 = (1-\kappa)^2 + \hat{\omega}^2 - \mu^{-1}$ to get an explicit form.

Moreover, the Jacobi map can be linked to another important matrix in the study of light propagation. Indeed, the Eq. \eqref{EqEvolJAB} can be rewritten as two $1^{\rm st}$ order differential equations\,:
\bea
\label{eq:deformation}
\frac{\di J^A_B}{\di \lambda}=\Scal^A_C J^C_B ~~~,~~~ \frac{\di \Scal^A_B}{\di \lambda}+\Scal^A_C \Scal^C_B=R^A_B ~~~,
\eea
where
\beq
\label{eq:Newdeformation}
\Scal^A_B\equiv \frac{\di J^A_C}{\di \lambda} (J^{-1})^C_B
\eeq
is called the \emph{deformation matrix}. This matrix can be decomposed as it is done with $\Acal$\,:
\beq
\label{eq:DeformationDecomposition}
\Scal^A_B=\hat\theta\,\delta^A_B+\left(\begin{array}{cc}
\hat\sigma_1&\hat\sigma_2\\
\hat\sigma_2&-\hat\sigma_1
\end{array}\right) ~~.
\eeq
As mentioned in \cite{Clarkson:2011br}, the symmetry of $\Scal$ is related to the fact that its antisymmetric part is proportional to $\nabla_{[\mu}k_{\nu]}$.
In such a way, the second equation of Eq. \eqref{eq:deformation} can be decomposed into the so-called \emph{Sachs equations} by considering respectively its trace and its trace-free parts\,:
\bea
\label{eq:SachsEq1}
& \frac{\di \hat\theta}{\di \lambda}+|\hat\sigma|^2+\hat\theta^2 = \frac12 \text{tr} R^A_B \equiv \Phi_{00} ~~, \\
\label{eq:SachsEq2}
& \frac{\di \hat\sigma}{\di \lambda}+2\hat\theta \hat\sigma \equiv \Psi_0 ~~,
\eea
where we have taken into account that $\text{tr}\left( \Scal^A_C\Scal^C_B \right)=2\left( \hat\theta^2+|\hat\sigma|^2 \right)$, used $\hat\sigma \equiv \hat\sigma_1 + i \hat\sigma_2$, and introduced $\Phi_{00}$ and $\Psi_0$ which are respectively called \textit{Ricci focusing} and \textit{Weyl focusing} (and whose definitions are given right after).

As we will see in the next section, quantities involved in Eqs. \eqref{eq:SachsEq1} and \eqref{eq:SachsEq2} can be derived in different ways. One of these, in particular, is by using direct expressions of the \emph{optical scalars} \cite{Clarkson:2011br}, namely\,:
\bea
\label{eq:OpticalScalarsDef}
\hat\theta\equiv\frac{1}{2}\nabla_\mu k^\mu ~~~ \mbox{(expansion scalar)} ~~, \\
\label{eq:OpticalScalarsDef2}
|\hat\sigma|^2\equiv\frac{1}{2}\nabla_\mu k_\nu\nabla^\mu k^\nu-\hat\theta^2 ~~~ \mbox{(shear scalar)} ~~.
\eea
Another way is to derive them from the so-called \emph{optical tidal matrix} $R^A_B$ once we have decomposed it as\,:
\beq
\label{Phi00Psi0}
R^A_B=\Phi_{00}\,\delta^A_B+\left( \begin{matrix} \text{Re}\Psi_0 & \text{Im}\Psi_0\\ \text{Im}\Psi_0 & -\text{Re}\Psi_0 \end{matrix} \right) ~~.
\eeq
By considering the well-known relation between the Riemann and the Weyl tensors\,:
\beq
C_{\alpha\beta\mu\nu} \equiv R_{\alpha\beta\mu\nu}-g_{\alpha[\mu}R_{\nu]\beta}+g_{\beta[\mu}R_{\nu]\alpha}+\frac{1}{3}R \, g_{\alpha[\mu}g_{\nu]\beta} ~~,
\eeq
we get that
\beq
\label{RABfromCurvatures}
R^A_B \equiv R_{\alpha\beta\mu\nu}k^\alpha k^\mu s^\beta_A s^\nu_B = -\frac{1}{2}R_{\mu\alpha}k^\alpha k^\mu\,\delta^A_B+C_{\alpha\beta\mu\nu}k^\alpha k^\mu s^\beta_A s^\nu_B ~~,
\eeq
due to the properties of the Sachs basis, Eq. \eqref{eq:Sachs}, and the condition $k^\alpha k_\alpha =0$. Therefore, by identification of Eqs. \rref{Phi00Psi0} and \rref{RABfromCurvatures}, we can write\,:
\beq
\label{eq:focusing}
\Phi_{00}=-\frac{1}{2}R_{\alpha\beta} k^\alpha k^\beta ~~~,~~~ \Psi_0=\frac{1}{2}C_{\alpha\beta\mu\nu}k^\alpha k^\mu \Sigma^\beta \Sigma^\nu ~~,
\eeq
where $\Sigma^\mu \equiv s^\mu_1+is^\mu_2$. Thanks to the Einstein equations, we can directly link the so-called \emph{Ricci focusing} $\Phi_{00}$ to the matter content (and, in particular, to its shape) because $k_\mu k^\mu = 0$ and therefore $R_{\mu\nu}k^\mu k^\nu=8\pi G\,T_{\mu\nu}k^\mu k^\nu$.
In the next section, we will evaluate all these quantities in a particular choice of coordinates in which the Jacobi map is given in a non-perturbative (exact) way\,: the geodesic light-cone gauge.

\section{Expressions in the Geodesic Light-cone gauge}
\label{Sec3}

We are going to present here the expression of the Jacobi map and the lensing quantities within the \emph{geodesic light-cone} (GLC) gauge. As the Jacobi map was obtained from the linear Jacobi equation, the (exact) results obtained here are valid within this approximation (of small angles). On the other hand, the GLC gauge is by construction assuming that no caustics form along the past light-cone (otherwise coordinate transformations break down) and can hence be used within the Born approximation. We will see that the GLC formalism has the advantage of computing lensing observables in a simple way, which is not the case in general (non-trivial) geometries.

As shown in \cite{Fanizza:2013doa}, the Jacobi map takes an exact form within the geodesic light-cone gauge \cite{P1}. This gauge consists of a timelike coordinate $\tau$ (identified with the proper time of the synchronous gauge), a null coordinate $w$ and two angles $\tilde\theta^a$. Its line element reads\,:
\beq
\label{GLCmetric}
\di s_{GLC}^2 = \Ups^2 \di w^2 - 2\Ups \di w \di \tau+ \gamma_{ab}(\di \ti{\theta}^a - U^a \di w)(\di \ti{\theta}^b - U^b \di w) ~~,
\eeq
where $\Ups$, $\gamma_{ab}$ and $U^a$ ($a,b \in \{ 1,2 \}$) are six free functions depending on all the coordinates.
In this framework, the zweibeins are written as $s^\mu_A=(s^\tau_A,0,s^a_A)$ (or equivalently $s_\mu^A=(0,s_w^A,s_a^A)$), $k^\mu =\omega \Ups^{-1}\delta^\mu_\tau$ (with $\omega$ a pure constant that can be chosen at will) and the solution of the Jacobi map equation, Eq. \rref{EqEvolJAB}, was derived in \cite{Fanizza:2013doa} as\,:
\beq
\label{JABandCaBOld}
J^A_B(\lambda,\lambda_o) = s_a^A(\lambda) \, C^a_B ~~,
\eeq
where $C^a_B$ is a constant matrix that we fix thanks to the initial conditions of Eq. \eqref{eq:initialConditions}. Hence we have that\,:
\beq
\frac{\di}{\di \lambda}J^A_B(\lambda_o,\lambda_o) = \left(k^\mu\partial_\mu s^A_a\right)_o C^a_B = \left(k^\tau\partial_\tau s^A_a\right)_o C^a_B = \left(\frac{k^\tau}{2}s^c_A\dot\gamma_{ca}\right)_o C^a_B ~~,
\eeq
where $(\ldots)^{\tdev} \equiv \partial_\tau (\ldots)$ and we used the parallel transport condition for the zweibeins, resulting from the last condition of Eq. \eqref{eq:Sachs}, and the Christoffel symbol $\Gamma_{b\tau}^a=\frac{1}{2}\gamma^{ac}\dot\gamma_{cb}$\,:
\beq
\label{ParallelTransport}
k^\mu\nabla_\mu s^a_A=0 ~~ \Rightarrow ~~ \dot s^a_A = - \Gamma_{\tau b}^a s^b_A = - \frac12 \gamma^{ac} \dot{\gamma}_{cb} s^b_A ~~.
\eeq
In this way, due to the second equation of Eq. \eqref{eq:initialConditions}, we get $\left(s^c_A\dot\gamma_{ca}\right)_oC^a_B = 2u_{\tau_o} \delta^A_B$ and by defining $\epsilon^{ab}$ as the pure antisymmetric symbol such that the inverse matrix of $\dot\gamma_{ab}$ is equal to $- \epsilon^{ab}\dot\gamma_{bc}\,\epsilon^{cd} / \det^{ab}\dot\gamma_{ab}$, we obtain an expression in which the zweibeins contribution is factorised\,:
\beq
C^a_B = - \left(2u_\tau\,\frac{\epsilon^{ab}\dot\gamma_{bc}\,\epsilon^{cd}}{\det^{ab}\dot\gamma_{ab}}s_d^B\right)_o ~~.
\eeq
Therefore, the Jacobi map appears as\,:
\beq
\label{JABandCaB}
J^A_B(\lambda,\lambda_o) = s_a^A(\lambda)\,\left[ - 2u_{\tau}\frac{\epsilon^{ac}\dot\gamma_{cd}\,\epsilon^{db}}{\det^{ab}\dot\gamma_{ab}} \right]_os^B_b(\lambda_o)\equiv s^A_a\left( \lambda \right)\Delta^{ab}\left( \lambda_o \right)s^B_b\left( \lambda_o \right) ~~.
\eeq

Having this expression at our disposal, one can now come back on the quantities defined in Sec. \ref{Sec2} in order to give their expression within the GLC gauge. The first of these quantities, the angular distance, is given by Eq. \rref{dAJAB} and depends only on the determinants $\gamma \equiv \det\gamma_{ab}$ and ${\det}^{ab}\dot{\gamma}_{ab}$\,:
\beq
\label{dAInGLC}
d_A = (\gamma \gamma_o)^{1/4} \, \sqrt{\det \Delta_o^{ab}} = \frac{2 u_{\tau_o}}{\sqrt{[\det^{ab} \dot{\gamma}_{ab}]_{o}}} (\gamma \gamma_o)^{1/4} ~~.
\eeq
In the same spirit, one can directly derive the expression of the magnification $\mu$ given in Eq. \rref{ExpressionsLensingQuantities}. This expression becomes within the GLC gauge\,:
\beq
\label{MagInGLC}
\mu = \frac{u_{\tau_o}^{-2}\bar{d}_A^{\,2}}{4\sqrt{\gamma \gamma_o}} \left[ {\det}^{ab} \dot{\gamma}_{ab} \right]_{o} \equiv \left(\frac{\bar d_A}{d_A}\right)^2 ~~.
\eeq
Let us notice the appearance in $\mu$, as for the other lensing quantities of Eq. \rref{ExpressionsLensingQuantities}, of the angular distance in the flat homogeneous and isotropic case $\bar{d}_A \equiv \bar d_A(\lambda_s)$. We hence need the expression of $\bar{d}_A$ in order to obtain $\mu$ solely in terms of GLC quantities. This expression is given by Eq. \rref{dAInGLC} and can be explicitly written in the GLC gauge as $\bar{d}_A = a^2(\tau) r^2$ with $a(\tau)$ the scale factor and $r = w - \int a^{-1}(\tau) \di \tau$ the conformal radius measured from the observer (see later the derivations of Appendix \ref{SecAppendixC}).

We can now deal with the other lensing quantities presented in Eq. \rref{ExpressionsLensingQuantities}, namely the convergence, shear and vorticity. These quantities all involve $\bar{d}_A$ (defined above) multiplying some combination of the Jacobi map components. According to Eq. \rref{JABandCaB}, we thus have to find the general expression of the zweibeins $s_a^A(\lambda)$ within the GLC gauge to express them. This resolution is presented in details in Appendix \ref{SecAppendixA} and we can see that the zweibeins then depend on an arbitrary angle $\beta$ related to their rotation freedom in the parallel transport condition.

Nevertheless, instead of presenting these general expressions here, one can use a combination of the lensing quantities that does not depend on this angle $\beta$ but only on notions already proven (and rather give the explicit formulas in Appendix \ref{SecAppendixB}). In fact, let us consider Eq. \rref{JABandCaB} and introduce $\Delta_o^{ab} \equiv \Delta^{ab}(\lambda_o)$. The convergence squared given by Eq. \rref{ExpressionsLensingQuantities} then reads\,:
\bea
(1-\kappa)^2&=&\frac{1}{4 \bar{d_A}^2}\left( J^1_1+J^2_2 \right)^2=\frac{1}{4 \bar{d_A}^2}\left( (J^1_1)^2+(J^2_2)^2+2J^1_1J^2_2 \right)\nonumber\\
&=&\frac{1}{4 \bar{d_A}^2}\left[ s^1_a\Delta^{ab}_os^1_b(\lambda_o)s^1_c\Delta^{cd}_os^1_d(\lambda_o) +s^2_a\Delta^{ab}_os^2_b(\lambda_o)s^2_c\Delta^{cd}_os^2_d(\lambda_o) +2 s^1_a\Delta^{ab}_os^1_b(\lambda_o)s^2_c\Delta^{cd}_os^2_d(\lambda_o) \right]\nonumber\\
&=&\frac{1}{4\bar{d_A}^2}\Delta^{ab}_o\Delta^{cd}_o\left[ s^1_as^1_c\left( s^1_bs^1_d \right)_o+s^2_as^2_c\left( s^2_bs^2_d \right)_o+2s^1_as^2_c\left( s^1_bs^2_d \right)_o \right] ~~.
\eea
In the same way, using Eqs. \rref{MatrixA}, \rref{ExpressionsLensingQuantities} and \rref{JABandCaB}, we find that\,:
\bea
\hat\omega^2&=&\frac{1}{4\bar{d_A}^2}\left( J^1_2-J^2_1\right)^2=\frac{1}{4\bar{d_A}^2}\Delta^{ab}_o\Delta^{cd}_o\left[ s^1_as^1_c\left( s^2_bs^2_d \right)_o+s^2_as^2_c\left( s^1_bs^1_d \right)_o-2s^1_as^2_c\left( s^2_bs^1_d \right)_o \right] ~~, \nonumber\\
\hat{\gamma}_1^2&=&\frac{1}{4\bar{d_A}^2}\left( J^1_1-J^2_2 \right)^2=\frac{1}{4 \bar{d_A}^2}\Delta^{ab}_o\Delta^{cd}_o\left[ s^1_as^1_c\left( s^1_bs^1_d \right)_o+s^2_as^2_c\left( s^2_bs^2_d \right)_o-2s^1_as^2_c\left( s^1_bs^2_d \right)_o \right] ~~, \nonumber\\
\hat{\gamma}_2^2&=&\frac{1}{4\bar{d_A}^2}\left( J^1_2+J^2_1 \right)^2=\frac{1}{4 \bar{d_A}^2}\Delta^{ab}_o\Delta^{cd}_o\left[ s^1_as^1_c\left( s^2_bs^2_d \right)_o+s^2_as^2_c\left( s^1_bs^1_d \right)_o+2s^1_as^2_c\left( s^2_bs^1_d \right)_o \right] ~~,
\eea
implying\,:
\bea
\left( 1-\kappa \right)^2+\hat\omega^2&=&\frac{1}{4\bar{d_A}^2}\Delta^{ab}_o\Delta^{cd}_o\left[ s^A_as^A_c\left( s^B_bs^B_d \right)_o+2s^1_as^2_c\left( \epsilon_{AB}s^A_bs^B_d \right)_o \right] ~~, \nonumber\\
\hat\gamma_1^2+\hat\gamma_2^2&=&\frac{1}{4\bar{d_A}^2}\Delta^{ab}_o\Delta^{cd}_o\left[ s^A_as^A_c\left( s^B_bs^B_d \right)_o-2s^1_as^2_c\left( \epsilon_{AB}s^A_bs^B_d \right)_o \right] ~~,
\eea
which can be simplified thanks to the identities\,:
\beq
s^A_as^A_b=\gamma_{ab} ~~~~~,~~~~~ \epsilon_{AB}\,s^A_as^B_b=\sqrt{\gamma}\,\epsilon_{ab} ~~,
\eeq
into the new expressions\,:
\bea
\left( 1-\kappa \right)^2+\hat\omega^2&=&\frac{1}{4\bar{d_A}^2}\Delta^{ab}_o\Delta^{cd}_o\left[ \gamma_{ac}\left(\gamma_{bd}\right)_o+2\sqrt{\gamma_o}\,s^1_as^2_c\epsilon_{bd} \right] ~~, \nonumber\\
\hat\gamma_1^2+\hat\gamma_2^2&=&\frac{1}{4\bar{d_A}^2}\Delta^{ab}_o\Delta^{cd}_o\left[ \gamma_{ac}\left(\gamma_{bd}\right)_o-2\sqrt{\gamma_o}\,s^1_as^2_c\epsilon_{bd} \right] ~~.
\eea
We can finally use
\beq
\left(\Delta^{ab}\gamma_{bc}\,\Delta^{cd}\right)_o\gamma_{ad} = 4u_{\tau_o}^2\left( \frac{\gamma\,\dot\gamma_{ab}\gamma^{bc}\dot\gamma_{cd}}{(\det^{ab}\dot\gamma_{ab})^2} \right)_o\gamma\,\gamma^{ad}
~~~~~,~~~~~
s_a^1\Delta^{ab}_o\epsilon_{bc}\Delta^{cd}_os_d^2 = \frac{4u_{\tau_o}^2 \sqrt{\gamma}}{\left(\det^{ab}\dot\gamma_{ab}\right)_o} ~~~~~,
\eeq
to obtain\,:
\bea
\label{LensingCombinationsInGLC}
\left( 1-\kappa \right)^2+\hat\omega^2&=&\left( \frac{u_{\tau_o}}{\bar d_A} \right)^2\left\{ \left[ \frac{\gamma\,\dot\gamma_{ab}\gamma^{bc}\dot\gamma_{cd}}{\left(\det^{ab}\dot\gamma_{ab}\right)^2} \right]_o\gamma\,\gamma^{ad}+2\frac{\sqrt{\gamma\,\gamma_o}}{\left( \det^{ab}\dot\gamma_{ab} \right)_o} \right\} ~~, \nonumber\\
\hat\gamma_1^2+\hat\gamma_2^2&=&\left( \frac{u_{\tau_o}}{\bar d_A} \right)^2\left\{ \left[ \frac{\gamma\,\dot\gamma_{ab}\gamma^{bc}\dot\gamma_{cd}}{\left(\det^{ab}\dot\gamma_{ab}\right)^2} \right]_o\gamma\,\gamma^{ad}-2\frac{\sqrt{\gamma\,\gamma_o}}{\left( \det^{ab}\dot\gamma_{ab} \right)_o} \right\} ~~.
\eea
One can then re-obtain the expression of the magnification in terms of the GLC quantities, Eq. \rref{MagInGLC}, by using Eqs. \rref{LensingCombinationsInGLC} and $\mu \equiv \left[(1 - \kappa)^2 + \hat\omega^2 - |\hat \gamma |^2 \right]^{-1}$.
Let us emphasize that the ratio $\bar d_A / u_{\tau_o}$ appears in all the quantities of $\mathcal A$ and because lensing does not depend on the observer's motion we choose our background distance $\bar d_A$ with the same observer motion $\bar u_{\tau_o}$ as the perturbed one, i.e. $\bar u_{\tau_o}=u_{\tau_0}$. In such a way, we can use\,:
\beq
\label{dAbar}
\left(\frac{\bar d_A}{\bar{u}_{\tau_o}}\right)^2=\sqrt{\bar\gamma}\left[ \frac{4\sqrt{\bar \gamma}}{\det\dot{\bar\gamma}_{ab}} \right]_o ~~,
\eeq
where we used $\overline{(\ldots)}$ to label background quantities.

Let us now express the deformation matrix and its elements, the optical scalars, in the GLC gauge. By its definition of Eq. \rref{eq:Newdeformation} this matrix is given in terms of the zweibeins by\,:
\beq
\label{eq:Def0}
\Scal^A_B=\frac{\di s^A_a}{\di \lambda} s^a_B = \frac{\omega}{2\Ups} s^a_A s^b_B\,\dot \gamma_{ab} ~~.
\eeq
Let us note that the observer terms of Eq. \rref{JABandCaB} have disappeared from this expression and that $\Scal^A_B$ is symmetric.
Using Eqs. \rref{eq:Def0}, the property $s^a_A s^b_A=\gamma^{ab}$, and the decomposition of Eq. \rref{eq:DeformationDecomposition}, we get the components of $\Scal^A_B$ (the optical scalars)\,:
\bea
\label{ThetaHat}
\hat\theta &=& \frac{\text{tr}\Scal^A_B}{2}=\omega\,\frac{\gamma^{ab}\dot\gamma_{ab}}{4\Ups}=\frac{\omega}{4\Ups} \frac{\dot\gamma}{\gamma} ~~, \\
\label{SigmaHat}
|\hat\sigma|^2 &=& \hat\sigma_1^2+\hat\sigma_2^2=\left( \frac{\text{tr}\Scal^A_B}{2} \right)^2-\det\Scal^A_B=\left(\frac{\omega}{4 \Ups} \frac{\dot\gamma}{\gamma}\right)^2 - \frac{\omega^2}{4\Ups^2}\,\frac{\det\dot\gamma_{ab}}{\gamma} ~~.
\eea
Hence the optical scalars are independent from $\beta$ as expected (as $\beta$ describes a $U(1)$-rotation freedom).
These expressions perfectly agree with the general definitions of Eqs. \eqref{eq:OpticalScalarsDef} and \rref{eq:OpticalScalarsDef2} with the usual GLC condition $k^\mu =\omega \Ups^{-1}\delta^\mu_\tau$.

Using the relations presented in Eq. \rref{eq:focusing}, or in both equivalent ways from Eqs. \rref{eq:SachsEq1}, \rref{eq:SachsEq2} or Eqs. \rref{eq:deformation}, \rref{eq:Def0}, we can also get the Ricci and Weyl focusing in the GLC gauge. We find\,:
\bea
\label{Phi00}
\Phi_{00}&=&\frac{\omega^2}{4\Ups^2} \left[ \gamma^{ab}\ddot \gamma_{ab} - \frac{\dot \Ups}{\Ups} \gamma^{ab}\dot\gamma_{ab} - \frac{1}{2}\gamma^{ab}\dot\gamma_{ac}\gamma^{cd}\dot\gamma_{db} \right] ~~, \nonumber \\
\text{Re}\Psi_0 &=& \frac{\omega^2}{4\Ups^2} \left[ \ddot \gamma_{ab} - \frac{\dot \Ups}{\Ups} \dot\gamma_{ab} - \frac{1}{2}\dot\gamma_{ac}\gamma^{cd}\dot\gamma_{db} \right]\left( s^a_1s^b_1-s^a_2s^b_2 \right) ~~, \nonumber\\
\text{Im}\Psi_0 &=& \frac{\omega^2}{4\Ups^2} \left[ \ddot \gamma_{ab} - \frac{\dot \Ups}{\Ups} \dot\gamma_{ab} - \frac{1}{2}\dot\gamma_{ac}\gamma^{cd}\dot\gamma_{db} \right]\left( s^a_1s^b_2+s^a_2s^b_1 \right) ~~,
\eea
where we have used for $\Phi_{00}$ that $s^a_A \, s^b_A = \gamma^{ab}$ and the full GLC expression for $\text{Re}\Psi_0$ and $\text{Im}\Psi_0$ can be obtained using the general expressions of the zweibeins in the GLC gauge (see Appendix \ref{SecAppendixA}).
In order to evaluate the modulus of $\Psi_0$, let us notice that, after some algebraic manipulations, we have\,:
\beq
(s^a_1 s^b_1-s^a_2 s^b_2)(s^c_1 s^d_1-s^c_2 s^d_2)+(s^a_1 s^b_2+s^a_2 s^b_1)(s^c_1 s^d_2+s^c_2 s^d_1)=\gamma^{ac}\gamma^{bd}+\gamma^{ad}\gamma^{bc}-\gamma^{ab}\gamma^{cd} ~~,
\eeq
so\,:
\beq
\label{Psi0}
|\Psi_0|^2=\frac{\omega^4}{16\Upsilon^4} \left[ \ddot \gamma_{ab} - \frac{\dot \Ups}{\Ups} \dot\gamma_{ab} - \frac{1}{2}\dot\gamma_{ae}\gamma^{ef}\dot\gamma_{fb} \right] \left[ \ddot \gamma_{cd} - \frac{\dot \Ups}{\Ups} \dot\gamma_{cd} - \frac{1}{2}\dot\gamma_{cg}\gamma^{gh}\dot\gamma_{hd} \right]\left( \gamma^{ac}\gamma^{bd}+\gamma^{ad}\gamma^{bc}-\gamma^{ab}\gamma^{cd} \right) ~~.
\eeq
This Weyl focusing term is then fully expressed in terms of the GLC metric elements, showing that only the combination $|\Psi_0|^2 \equiv (\text{Re}\Psi_0)^2 + (\text{Im}\Psi_0)^2$ is independent from the angle $\beta$. The pure constant $\omega$ can be chosen at will, e.g. one can take for simplicity $\omega = 1$. The lensing quantities of this section take a very simple interpretation in the homogeneous and isotropic context, as shown in Appendix \ref{SecAppendixC}. One could also have chosen another decomposition of the amplification matrix as the product of a rotation matrix and a symmetric matrix \cite{schneider1999gravitational,Fleury:2014gha}. We preferred to keep the simplest convention as the latter can be easily derived from quantities presented here. In any case, the application of the next section will contain no vorticity and will hence be independent from these choices in decomposition.

Finally, one should recall that our GLC approach and the expressions presented here are general (again, as long as no caustic is formed). Hence we conclude that the geodesic light-cone gauge is perfectly adapted to the computation of lensing quantities and we can indeed appreciate the great simplicity of these results compared to what is expected in other general geometries. The literature on lensing is well furnished with more mathematical approaches (see e.g. \cite{Frittelli:2000bb,Frittelli:2000bc,petters2001singularity}) but we can argue that the GLC coordinates and the expressions here derived are easier to deal with as they are more explicit. They can be directly applied, in principle, to a large number of geometries through coordinate transformations. The next section will be dedicated to a useful application and a good illustration of this last point. In particular, we will evaluate the lensing quantities in the well-known geometrical scenario of the LTB metric.

\section{Off-center observer in LTB coordinates - Angular distance and Amplification matrix}
\label{Sec4}

We are now going to express our results in terms of the \emph{Lema\^itre-Tolman-Bondi} (LTB) coordinates defined by the line element (see e.g. \cite{2012EPJC...72.2242R,2012CRPhy..13..682C,2014arXiv1408.4442Z,2014PhRvD..90j3510N} for recent applications)\,:
\beq
\di s^2=-\di t^2+X^2(t,r) \di r^2+A^2(t,r)\left[ \di \theta^2+\sin^2\theta\, \di \phi^2 \right].
\eeq
As well-known, this metric shows a symmetry for any 3D rotation around its center at $r=0$. Therefore any observer located at that position cannot detect any anisotropy but only radial inhomogeneity. Hence the center of coordinates appears as a preferred point with respect to the other ones. In such a way, whoever is located far from the center measures also anisotropy along the axis connecting him/herself with the center $r=0$. Having this in mind, we want to describe the quantities we derived in the last section as seen by an off-center observer within an LTB scenario. So we underline that the centers of the GLC and the LTB coordinates are displaced by a given distance $d$ and the observer sees an inhomogeneous as well as anisotropic spacetime.

\begin{figure}[ht!]
\centering
\begin{tikzpicture}
  \node [xshift=2cm] at (current page.south east)
    {\includegraphics[width=7cm]{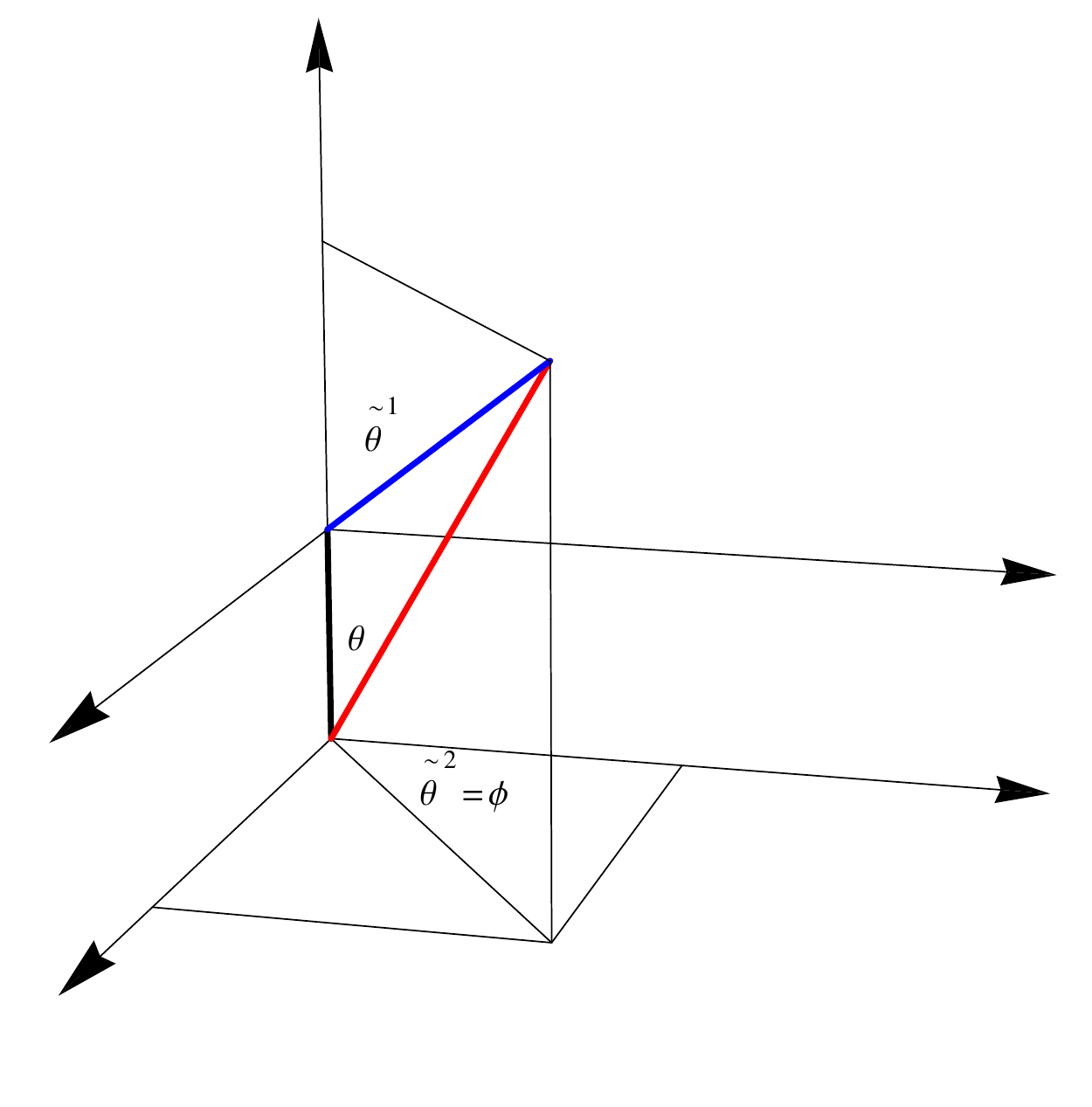}};      
  \node [xshift=9cm] at (current page.south east)
    {\includegraphics[width=3cm]{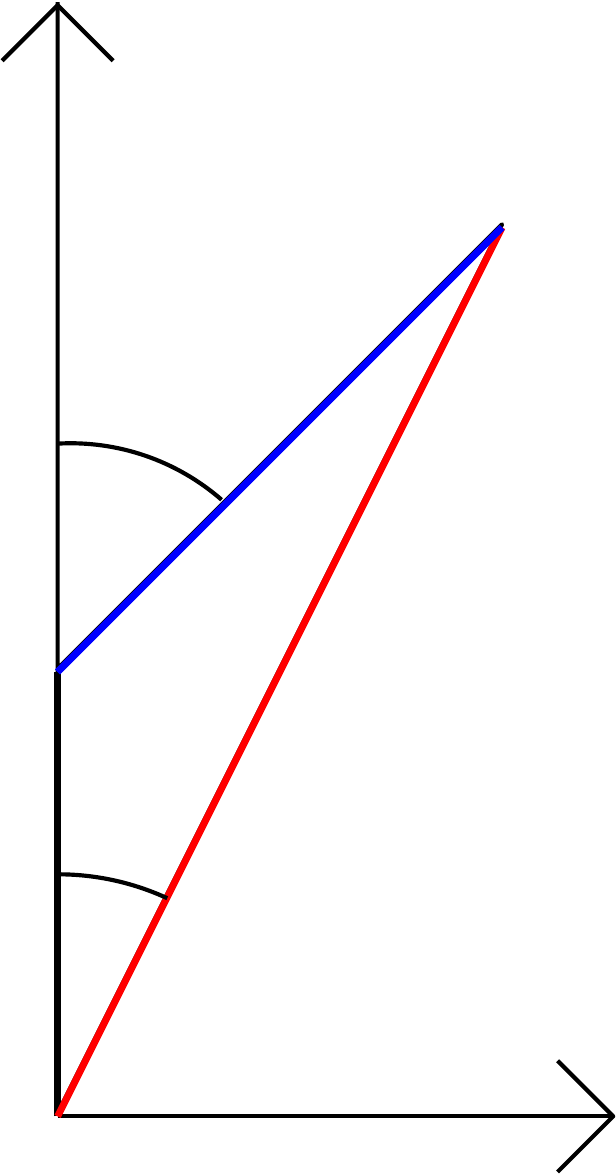}};

% Left picture
\node[anchor=west] at (-1.9,-3) (description) {$x$};
\node[anchor=west] at (5.5,-1.7) (description) {$y$};

\node[anchor=west] at (-1.9,-1.5) (description) {$x$};
\node[anchor=west] at (5.5,-0.2) (description) {$y$};
\node[anchor=west] at (0.35,3.7) (description) {$z$};

\node[anchor=west] at (0.1,-0.65) (description) {$\tbf{d}$};

% Right picture
\node[anchor=west] at (7.5,3.5) (description) {$z$};
\node[anchor=west] at (10.5,-2.5) (description) {$(x,y)$-plane};

\node[anchor=west] at (7.2,-1.5) (description) {$\tbf{d}$};
\node[anchor=west] at (8.9,-0.6) (description) {$\textcolor{red}{r}$};
\node[anchor=west] at (8.5,0.1) (description) {$\textcolor{blue}{\tilde r}$};

\node[anchor=west] at (8.0,1.5) (description) {$\tilde{\theta}^1 \equiv \tilde{\theta}$};
\node[anchor=west] at (8,-0.7) (description) {$\theta$};

\end{tikzpicture}
\centering
\caption{Coordinates relation among the LTB radius $r$ (red) and the observer one $\tilde r$ (blue). These relations are made by imposing that the azimuthal angles are equal ($\tilde\theta^2=\phi$). \emph{Left}\,: the 3D illustration ; \emph{Right}\,: within the $(\tilde{\theta}^2 \equiv \phi) = {\rm const}$ plane.}
\label{fig:Coordinates}
\end{figure}
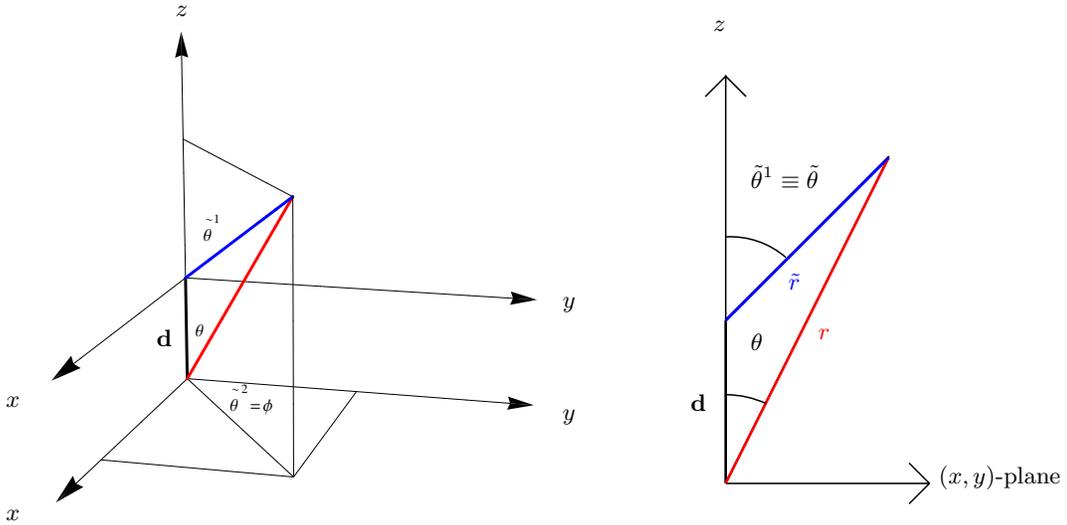

For a matter of simplicity, and with no loss of generality, let us require that the azimuthal angles in the two different coordinate systems are equal (see Fig. \ref{fig:Coordinates}). So, after some easy geometrical considerations, we get\,:
\bea
\label{eq:coordinatesTransformation}
\tau&=&t\nonumber\\
w&=&W(t,r,\theta)\nonumber\\
\tilde\theta^1&=&\arccos\left( \frac{r\cos\theta-d}{\sqrt{r^2+d^2-2 r d\cos\theta}} \right)\nonumber\\
\tilde\theta^2&=&\phi
\eea
where the identity $\tau=t$, proved in \cite{P2}, holds thanks to the synchronous gauge choice of the LTB metric. Here $W(t,r,\theta)$ is an implicit function that must obey the following relations $g^{ww}_{GLC}=g^{wa}_{GLC}=0$, i.e.\,:
\beq
\frac{\partial_\theta W}{\partial_t W}=\frac{A^2\,d\sin\theta}{\sqrt{A^2\,d^2\sin^2\theta+r^2X^2\left( r-d\cos\theta \right)^2}}\quad,\quad
\frac{\partial_r W}{\partial_t W}=\frac{r\left( r-d\cos\theta \right)X^2}{\sqrt{A^2\,d^2\sin^2\theta+r^2X^2\left( r-d\cos\theta \right)^2}} ~~.
\eeq

Moreover, the GLC metric elements are\,:
\beq
\label{eq:GLCvsLTB}
\Upsilon=\frac{1}{\partial_t W}\quad,\quad
U^a=\left( \begin{array}{cc} 0&0 \end{array} \right)\quad,\quad
\gamma^{ab}=\left( \begin{array}{cc} \frac{A^2\,d^2\sin^2\theta+r^2X^2\left( r-d\cos\theta \right)^2}{A^2 X^2\left( d^2+r^2-2rd\cos\theta \right)^2}&0\\0&A^{-2}\sin^{-2}\theta \end{array} \right) ~~,
\eeq
and the derivative with respect to $\tau$ of a generic function $f$ becomes\,:
\beq
\partial_\tau f=\partial_tf-\frac{r\left( r-d\cos\theta\right)\partial_rf+d\sin\theta\,\partial_\theta f}{\sqrt{A^2\,d^2\sin^2\theta+r^2 X^2\left( r-d\cos\theta \right)^2}} ~~.
\eeq
Following \cite{Fanizza:2013doa}, we are able to write the angular distance for a generic off-center observer in the LTB coordinates. This result has been already obtained and studied in the literature \cite{Blomqvist:2009ps,Cosmai:2013iga,2012GReGr..44.2449V,2007JCAP...02..013B,2007PhRvD..75b3506A,Fleury:2014gha, Bolejko:2012uj} but no explicit expressions (up to our knowledge) have been shown concerning lensing quantities. As we are going to show, our results for lensing quantities are explicitly given in terms of the LTB coordinates.

What we need for writing the angular distance is a source term given by\,:
\beq
\label{eq:source}
\sqrt{\gamma}=\frac{A^2 X \left( r^2+d^2-2\,rd\cos\theta \right)}{\sqrt{A^2\,d^2\sin^2\theta+r^2X^2\left( r-d\cos\theta \right)^2}}\sin\theta
\eeq
and an observer one (see Eq. \rref{dAInGLC}) which can be written as\,:
\beq
\label{eq:observer}
\left( \frac{4\sqrt{\gamma}}{\det^{ab}\dot\gamma_{ab}} \right)_o = \frac{A_0(d)}{d\,X_0(d)} \frac{G(\theta_o)}{\cos\frac{\theta_o}{2}} \equiv \frac{A_0(d)}{d\,X_0(d)} \frac{\tilde{G}(\tilde \theta)}{\sin \tilde \theta} ~~.
\eeq
Here we have considered that the observer position is at $t=t_o$, $r=d$, $(\theta,\phi)=(\theta_o,\phi_o)$ in the LTB coordinates and we made use of the third relation of Eqs. \eqref{eq:coordinatesTransformation} with $r=d$ to prove that $\cos(\theta_o/2) = \sin\tilde\theta$ ($\tilde\theta$ being by definition the angle seen by the observer in the GLC coordinates). The function $\tilde{G}(\tilde\theta)$ is a function of $\tilde\theta$ and the derivatives of $A(t,r)$ and $X(t,r)$ such that $\tilde{G}(\tilde\theta) \rightarrow 1$ when $\theta_o \rightarrow 0$. Hence, fixing the angular dependence of LTB coordinates such that the observer stands at $(\theta_o,\phi_o) \equiv (0,0)$, the observer term reduces to $(\sin\tilde\theta)^{-1}$ (in accordance with \cite{P1,P2,P3,P4,P5}) and Eqs. \rref{dAbar}, \eqref{eq:source}, \eqref{eq:observer} lead to\,:
\beq
\label{eq:Off-Center-dA}
d_A^2=\frac{A^2 X \left( r^2+d^2-2\,rd\cos\theta \right)}{\sqrt{A^2\,d^2\sin^2\theta+r^2X^2\left( r-d\cos\theta \right)^2}}\frac{A_0(d)}{d\,X_0(d)}\frac{\sin\theta}{\sin\tilde\theta} ~~.
\eeq
Let us add that in the LTB metric we can always manage a residual gauge degree of freedom in order to fix $A(t_o,r)\equiv A_0(r)=r$. Moreover, in the flat case, using the off-diagonal Einstein equations, we can write $X(t,r)=\partial_rA(t,r)$.

Before computing the lensing quantities, let us discuss the flat FLRW case in order to find the expression for $\bar d_A$. In this limit we have $A(t,r)\rightarrow r\,a( t)$ and $X(t,r)\rightarrow a(t)$, therefore Eq. \rref{eq:Off-Center-dA} becomes\,:
\beq
\label{dAbarUsed}
\bar d_A^2=\frac{a^2 \sqrt{r^2+d^2-2\,rd\cos\theta} ~ r\sin\theta}{\sin\tilde\theta} ~~.
\eeq
This expression appears very different from the usual one in the FLRW case (see Appendix \ref{SecAppendixC}). However, by noticing that we can re-express it in terms of the observer radial coordinate $\tilde r=\sqrt{r^2+d^2-2\,rd\cos\theta}$, we have that $r\sin\theta = \tilde r\sin\tilde\theta$ and we obtain $\bar d_A^2=\tilde r^2 a(t)^2$ just as expected.

Now we proceed with the evaluation of the amplification matrix in such a context. From Eq. \rref{eq:GLCvsLTB} we notice that $\gamma_{ab}$ is diagonal and it easily gives $\hat\omega=0$, i.e. no vorticity is present (as it can be directly obtained from Eq. \rref{eq:amplificationQuantities1}). Moreover, the remaining lensing quantities become in this LTB application\,:
\bea
\label{LensingQuantitiesInLTB}
\mu &=& \left( \bar d_A / d_A \right)^2 ~~, \nonumber\\
(1-\kappa)^2&=&\left( \frac{u_{\tau o}}{\bar d_A} \right)^2\left[ \gamma_{11}\left( \frac{\gamma_{11}}{(\dot{\gamma}_{11})^2} \right)_o+\gamma_{22}\left( \frac{\gamma_{22}}{(\dot{\gamma}_{22})^2} \right)_o+2\sqrt{\gamma_{11}\gamma_{22}}\left( \frac{\sqrt{\gamma_{11}\gamma_{22}}}{\dot\gamma_{11}\dot\gamma_{22}} \right)_o \right] ~~, \nonumber\\
|\hat\gamma|^2&=&\left( \frac{u_{\tau o}}{\bar d_A} \right)^2\left[ \gamma_{11}\left( \frac{\gamma_{11}}{(\dot{\gamma}_{11})^2} \right)_o+\gamma_{22}\left( \frac{\gamma_{22}}{(\dot{\gamma}_{22})^2} \right)_o-2\sqrt{\gamma_{11}\gamma_{22}}\left( \frac{\sqrt{\gamma_{11}\gamma_{22}}}{\dot\gamma_{11}\dot\gamma_{22}} \right)_o \right] ~~.
\eea
According to the fact that, within this framework, a static observer is also geodesic, we choose $u_{\tau_o}=1$. Using now the expression of $\bar d_A$ described in Eq. \rref{dAbarUsed} we get that\,:
\bea
\label{LensingQuantitiesExplicitInLTB}
\mu &=& \frac{r\,d\,X_0(d)\,a^2(t)}{A_0(d)\,A^2(t,r)\,X(t,r)}\sqrt{\frac{A^2(t,r)\,d^2\sin^2\theta+r^2X^2(t,r)\left( r-d\cos\theta \right)^2}{d^2+r^2-2 r d\cos\theta}} ~~, \nonumber\\
(1-\kappa)^2 &=& \frac{A^2(t,r)}{4\sin\theta\,d^2 r\,a^2(t) X^2_0(d) \sqrt{d^2-2 d r \cos\theta+r^2}} \Bigg[d^2 \sin ^2\theta X^2_0(d) \nonumber\\
&+& \frac{A^2_0(d) X^2(t,r) \left(d^2-2 d r \cos\theta+r^2\right)^2}{d^2 \sin ^2\theta A^2(t,r)+r^2 X^2(t,r) (r-d \cos\theta)^2}+\frac{2 d \sin\theta A_0(d) X_0(d) X(t,r) \left(d^2-2 d r \cos\theta+r^2\right)}{\sqrt{d^2 \sin ^2\theta A^2(t,r)+r^2 X^2(t,r) (r-d \cos\theta)^2}} \Bigg] ~~, \nonumber\\
|\hat\gamma|^2 &=& \frac{A^2(t,r)}{4\sin\theta\,d^2 r\,a^2(t) X^2_0(d) \sqrt{d^2-2 d r \cos\theta+r^2}} \Bigg[d^2 \sin ^2\theta X^2_0(d) \nonumber\\
&+& \frac{A^2_0(d) X^2(t,r) \left(d^2-2 d r \cos\theta+r^2\right)^2}{d^2 \sin ^2\theta A^2(t,r)+r^2 X^2(t,r) (r-d \cos\theta)^2}-\frac{2 d \sin\theta A_0(d) X_0(d) X(t,r) \left(d^2-2 d r \cos\theta+r^2\right)}{\sqrt{d^2 \sin ^2\theta A^2(t,r)+r^2 X^2(t,r) (r-d \cos\theta)^2}} \Bigg] ~~.
\eea
As expected from our choice of the observer angular position $\theta_o = 0$, we have expressions which are here independent (by symmetry) from the azimuthal angle $\phi$. Up to our knowledge these expressions of lensing quantities have not been shown in the literature.

In the following section we will concentrate our attention on the magnification and this due to its direct physical interpretation. In fact, we can relate the magnification to the luminosity flux $\Phi$ and write\,:
\beq
\mu=\left( \frac{\bar d_A}{d_A} \right)^2=\frac{(1+z)^4\bar d_A^2}{(1+z)^4 d_A^2}=\frac{\bar d_L^2}{d_L^2}=\frac{\Phi}{\bar\Phi} ~~,
\eeq
where we adopt the same observed redshift $z$ for both distances. This means that, at a given measured redshift, the magnification directly contains all the effects due to the LTB inhomogeneity. Therefore,  the greater the magnification, the closer (or equivalently the brighter) the source and for $\mu>1$ ($\mu<1$) objects are less (more) far than the homogeneous scenario predicts. We will discuss the magnification using two particular deviations from the homogeneous scenario\,: an LTB CDM model and an LTB $\Lambda$CDM one.

\section{Off-center observer in LTB coordinates - CDM and $\Lambda$CDM models}
\label{Sec5}

In this section, we will study the dynamics of the LTB models within general relativity and give the redshift evolution of some lensing quantities previously written. First of all, let us recall that for a generic collection of non-interacting perfect barotropic fluids we have the following relations linking the Hubble parameter to the LTB metric elements \cite{Enqvist:2006cg}\,:
\beq
\label{eq:LTBEE}
H^2(t,r)=H^2_0(r)\sum_n\Omega_{n0}( r)\left[ \frac{A_0(r)}{A(t,r)} \right]^{\alpha_n}\qquad,\qquad\sum_n\Omega_{n0}( r)=1\qquad,\qquad X(t,r)=\frac{\partial_r A(t,r)}{\sqrt{1-k(r)}} ~~,
\eeq
where $H(t,r)\equiv \partial_t A(t,r)/A(t,r)$, $H_0( r)\equiv H(t_0,r)$ is the inhomogeneous Hubble function evaluated ``today'' and $k( r)$ is a free-function that we can interpret as the inhomogeneous spatial curvature. $\Omega_{n0}(r) \equiv \Omega_{n}(t_0,r)$ and $\alpha_n$ are respectively the actual value of the density and the exponent for the evolution of the $n$-th fluid. From now on and for simplicity of our illustration we consider only a flat model with $k(r)=0$ (hence considering only the decaying mode). The non-zero curvature case (see e.g. \cite{2008PhRvD..78d3504Z,2012GReGr..44.2449V}) will be addressed in a future publication intended for more specific (and realistic) examples.

The mass $M_0(r)$ contained within a 3-dimensional sphere of radius $r$ is given by $M_0(r) = A^3_0(r) \Omega_{m0}(r) H^2_0(r) / 2 G$, where $G$ is the Cavendish (gravitational) constant. For such a reason, we can define the matter density today as $\rho_0(r) \equiv M_0(r) / \left( 4 \pi A^3_0(r) / 3 \right)$, which appears, from the second of Eqs. \eqref{eq:LTBEE}, as\,:
\beq
\label{RelDensityHo}
2\,G\rho_0( r)=\frac{3}{4 \pi}\Omega_{m0}(r)\,H^2_0(r)=\frac{3}{4 \pi}\left[ 1-\sum_{n\ne \text{matter}}\Omega_{n0}(r) \right]\,H^2_0( r) ~~,
\eeq
and where the time evolving density is given by $\rho(t,r)$ and satisfies $\rho(t,r) A(t,r)^3 = \rho_0(r) A_0(r)^3$.
Therefore, the total matter density is proportional to $H^2_0( r)$ by a function of $r$. In the following, we will consider two different solutions\,: the CDM case, with $\Omega_{m0}(r)=1$, and the $\Lambda$CDM one, where $\Omega_{m0}(r) = 1-\Omega_{\Lambda 0} (r)=1-\Omega_{\Lambda0}\left(\frac{H_0}{H_0(r)}\right)^2$. Let us add that here $\Omega_{\Lambda0}$ and $H_0$ represent the values of the homogeneous case respectively for the cosmological constant density and the Hubble constant and will here be interpreted as our background quantities (that we recover at $r \rightarrow \infty$). Hence $H_0 \equiv \lim_{r \rightarrow \infty} H_0(r)$ and $\Omega_{\Lambda 0}$ is such that $H^2_0(r) \Omega_{\Lambda 0}(r) = \Omega_{\Lambda0} H_0^2 \equiv \Lambda / 3$. Therefore, for our purpose, $H_0(r)$ completely takes into account the density profile of matter and so, by choosing it, we can directly study the under/overdensity we want to consider.

Let us now take an ansatz for $H_0(r)$ and model our inhomogeneity as follows\,:
\beq
\label{AnsatzH0}
H_0( r)=H_0\sqrt{1-\frac{H_0^2-H^2_\text{in}}{H^2_0}\frac{\tanh\left( \frac{d-r_0}{2\,\Delta r} \right)-\tanh\left( \frac{r-r_0}{2\,\Delta r}\right)}{\tanh\left( \frac{d-r_0}{2\,\Delta r}\right)+\tanh\left( \frac{r_0}{2\,\Delta r} \right)}}
\eeq
where $r_0$ is the radius of the under/overdensity, $d$ is the distance between the observer and the center of the inhomogeneity that appears in Eq. \eqref{eq:coordinatesTransformation}. This distance is assumed to be much longer than the under/overdensity size, i.e. $d \gg r_0$, in order to have $\lim_{r \rightarrow \infty} H_0(r) = H_0$ from Eq. \rref{AnsatzH0}. Here $\Delta r$ is the transition scale from the bubble to the background and it is assumed to be such that $\Delta r \ll r_o \ll d$ . Moreover, $H_0$ is the background value of the Hubble constant, i.e. $70 ~ \rm{km} \, \rm{s}^{-1} \, \rm{Mpc}^{-1}$, and $H_{in}$ is the Hubble constant at the center of the inhomogeneous region. Using the ansatz of Eq. \rref{AnsatzH0} into Eq. \rref{RelDensityHo} we express the density within the LTB coordinates and get that the background matter density will be proportional to $H^2_0$ while the one inside the bubble will be proportional to $H^2_{in}$ (for a sharp transition $\Delta r \ll r_o$, as assumed here). Therefore, modelling an under (over) density means choosing $H_{in}$ lower (greater) than $H_0$.

We can now discuss these general features in two particular models\,: the CDM and $\Lambda$CDM models. For that let us recall the definitions expressed in Eq. \rref{eq:LTBEE} and see that the time $t$ is given by \cite{Enqvist:2006cg}\,:
\beq
t_0 - t =\int_{A(t,r)}^{A_0(r)} \frac{\di A}{A H(t,r)} = \frac{1}{H_0(r)} \int_{A(t,r)/A_0(r)}^{1} \frac{\di x}{x \sqrt{\Omega_{m0}(r) x^{-3} + \Omega_{\Lambda 0}(r)}} ~~,
\eeq
in which we could add a curvature component $\Omega_{c 0}(r) x^{-2}$ but here we assumed $k(r) = 0$.
One can then inverse this relation in order to get $A(t,r)$ and we will assume hereafter, as we are authorised to do, that $A_0(r) = r$. The inversion of this relation gives us the following expressions for the expansion factor\,:
\begin{itemize}
\item{\textbf{inhomogeneous CDM model\,:} here the expansion factor can be described by the following solution\,:
\beq
A(t,r)=r\left[ 1+\frac{3}{2}\,H_0( r)\,t \right]^{2/3} ~~,
\eeq
with $\Omega_{m0}(r)=1$ (hence $\Omega_{\Lambda 0}(r) = 0$), and where we have chosen $t_0=0$. In such a context, the matter density is immediately given by $\rho_0( r) = 3 H^2_0( r)/ 8 \pi G$.}
\item{\textbf{inhomogeneous $\Lambda$CDM model\,:} here the expansion factor appears as
\beq
A(t,r)=r \left[ \frac{1-\Omega_{\Lambda0}( r)}{\Omega_{\Lambda0}( r)} \right]^{1/3}\left( \sinh \left[ \text{arcsinh}\sqrt{\frac{\Omega_{\Lambda0}( r)}{1-\Omega_{\Lambda0}( r)}}+\frac{3}{2}\sqrt{\Omega_{\Lambda0}( r)}\,H_0( r)\,t \right] \right)^{2/3} ~~,
\eeq
where $t_0=0$, $\Omega_{\Lambda0}( r)+\Omega_{m0}( r)=1$ and $H_0^2\,\Omega_{\Lambda0}=H^2_0( r)\,\Omega_{\Lambda0}( r)$. This last condition follows from the fact that $\Lambda$ is a constant. According to that, we can rewrite this solution in terms of the parameters of $H_0( r)$ and $\Omega_{\Lambda0}$, $H_0$, namely\,:
\beq
A(t,r)=r \left[ \frac{H^2_0( r)}{\Omega_{\Lambda0}\,H^2_0}-1 \right]^{1/3} \left( \sinh \left[ \text{arcsinh}\left(H_0\sqrt{\frac{\Omega_{\Lambda0}}{H^2_0( r)-\Omega_{\Lambda0}\,H^2_0}}\right)+\frac{3}{2}\sqrt{\Omega_{\Lambda0}}\,H_0\,t \right] \right)^{2/3} ~~.
\eeq}
We consider for these parameters the background values $H_0=70 ~ \rm{km} \, \rm{s}^{-1} \, \rm{Mpc}^{-1}$ and $\Omega_{\Lambda0}=0.68$. Moreover, the matter density is now given by $\rho_0(r)= 3 \left[H^2_0( r)-\Omega_{\Lambda0} H^2_0\right]/ 8 \pi G$. This means that, aside from a constant value, the dependence by the radius is the same as in the CDM model.
\end{itemize}

Thanks to our choices, the shape of the inhomogeneities is entirely given in both cases by the choice of the parameters $r_0$, $\Delta r$ and $H_\text{in}$ within $H_0( r)$ (see Eq. \rref{AnsatzH0}). As an illustrative example, we chose an under/overdensity located at $d=10$, $100$, or $1000$ Mpc from us, with a radius $r_0=1$ Mpc and a transition shell of $\Delta r=0.1$ Mpc in size. Moreover, for our purposes, we define a density contrast $\delta( r)$ with respect to the background density, $\rho_\text{BG} \equiv 3 \Omega_{m0} H_0^2 / 8 \pi G = 3 (1 - \Omega_{\Lambda 0}) H_0^2 / 8 \pi G$, as\,:
\beq
\delta( r)\equiv\frac{\rho_0( r)-\rho_{BG}}{\rho_{BG}}=\frac{H^2_0( r)-H^2_0}{H^2_0\left( 1-\Omega_{\Lambda0} \right)} ~~,
\eeq
which is valid for both models (by taking $\Omega_{\Lambda 0} = 0$ in the CDM case). In such a way, the maximum contrast will be given at the center of the bubble by $\delta_\text{max} = \delta(0) =H^2_\text{in}-H^2_0 / H^2_0\left( 1-\Omega_{\Lambda0} \right)$. Therefore, if we consider a variation of $2 ~ \rm{km} \, \rm{s}^{-1} \, \rm{Mpc}^{-1}$ for $H_0$ we have a maximum density contrast $\delta_{\text{max}}$ equal to $-0.056$ for the underdensity and to $0.058$ for the overdensity in the CDM scenario instead of $\delta_{\text{max}}$ which goes from $-0.176$ to $0.181$ respectively for under and over densities in the $\Lambda$CDM scenario. The numerical values chosen here are not representative of the biggest voids in the Universe for which $\delta < 0.8$ and $r_0 \sim 100$ Mpc, but our goal here is simply an illustration of our derivations.

\begin{figure}[ht!]
\centering

\subfigure[Overdensity at d=10 Mpc]{
	\includegraphics[height=3.8cm,width=7.5cm]{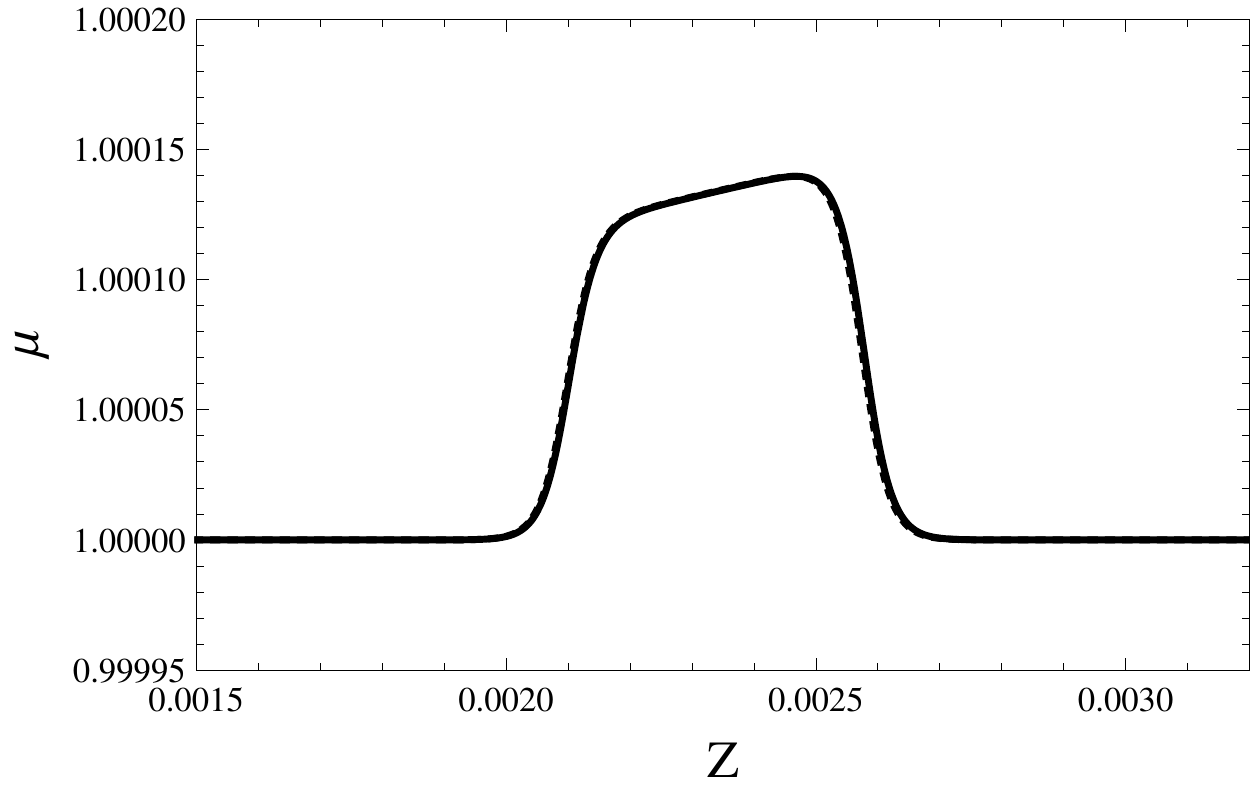}~~~~~
	\includegraphics[height=3.8cm,width=7.5cm]{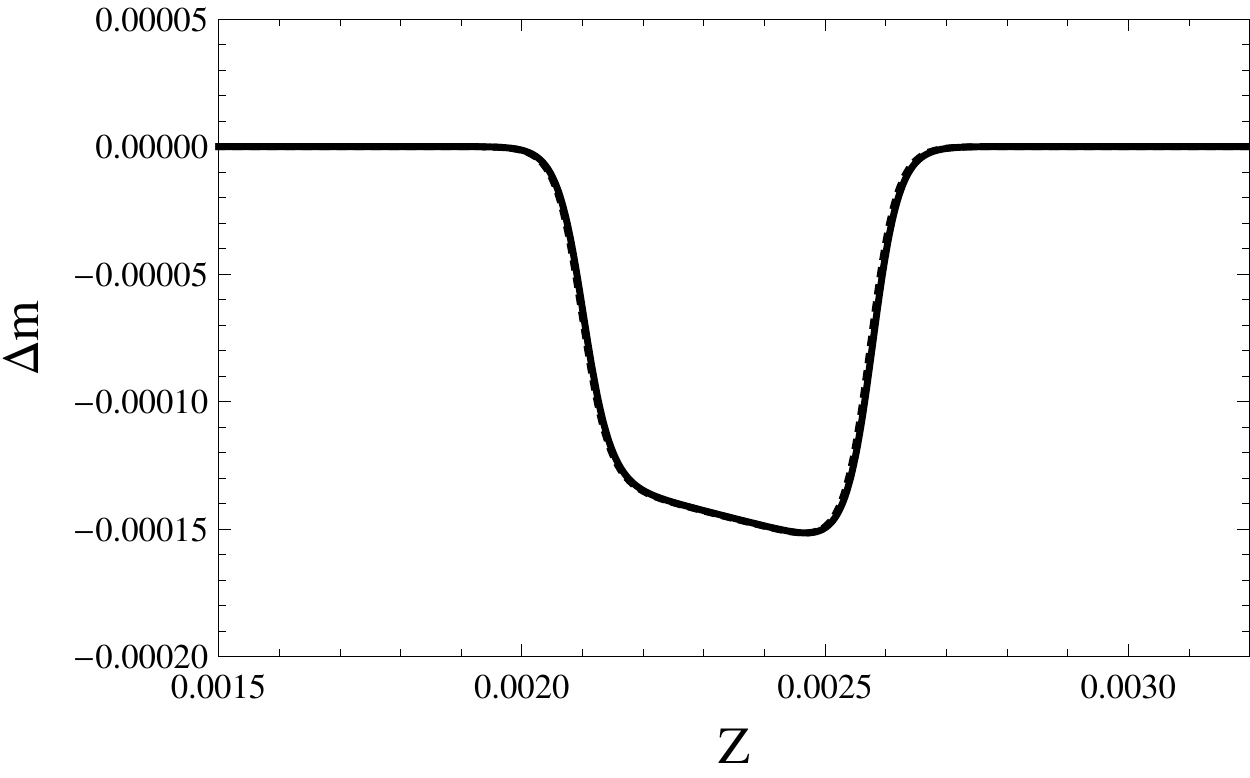}~~~~~~~~~~
	\label{fig:Over10Mpc}
 }

\subfigure[Underdensity at d=10 Mpc]{
 	\includegraphics[height=3.8cm,width=7.5cm]{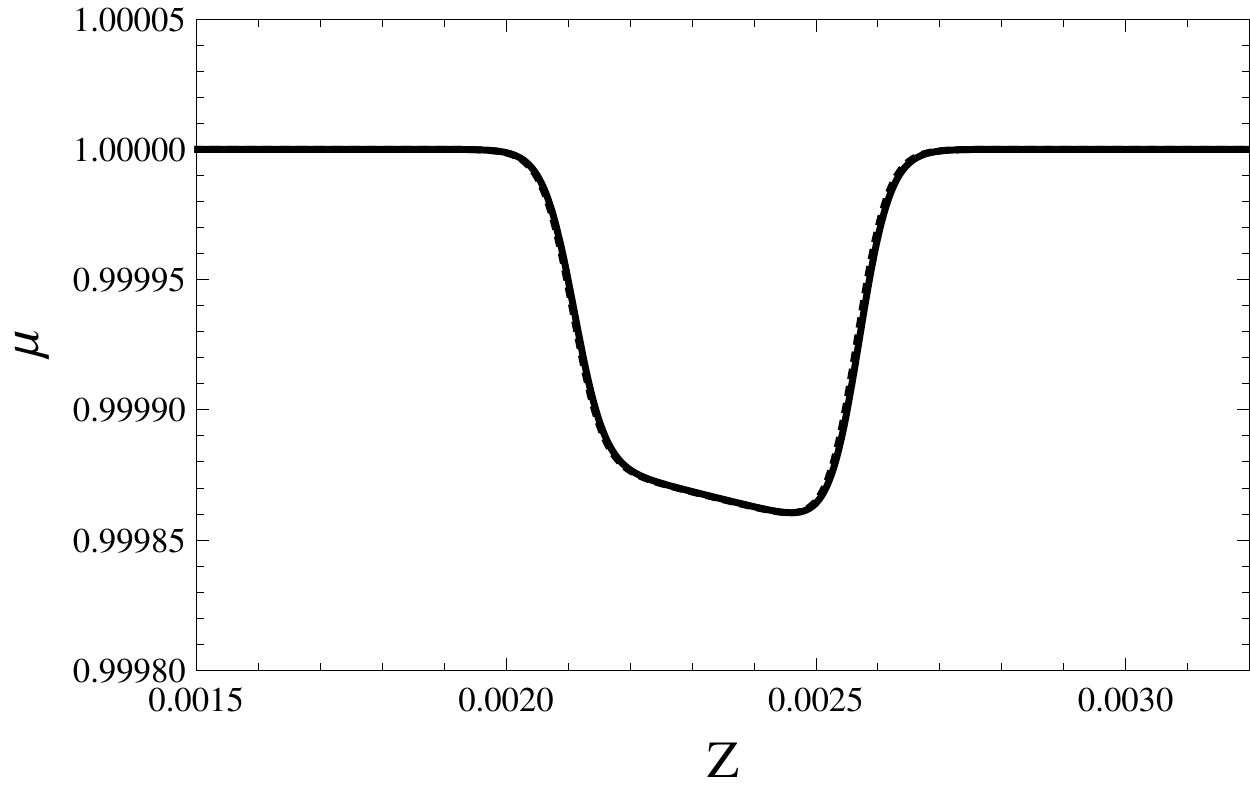}~~~~~
 	\includegraphics[height=3.8cm,width=7.5cm]{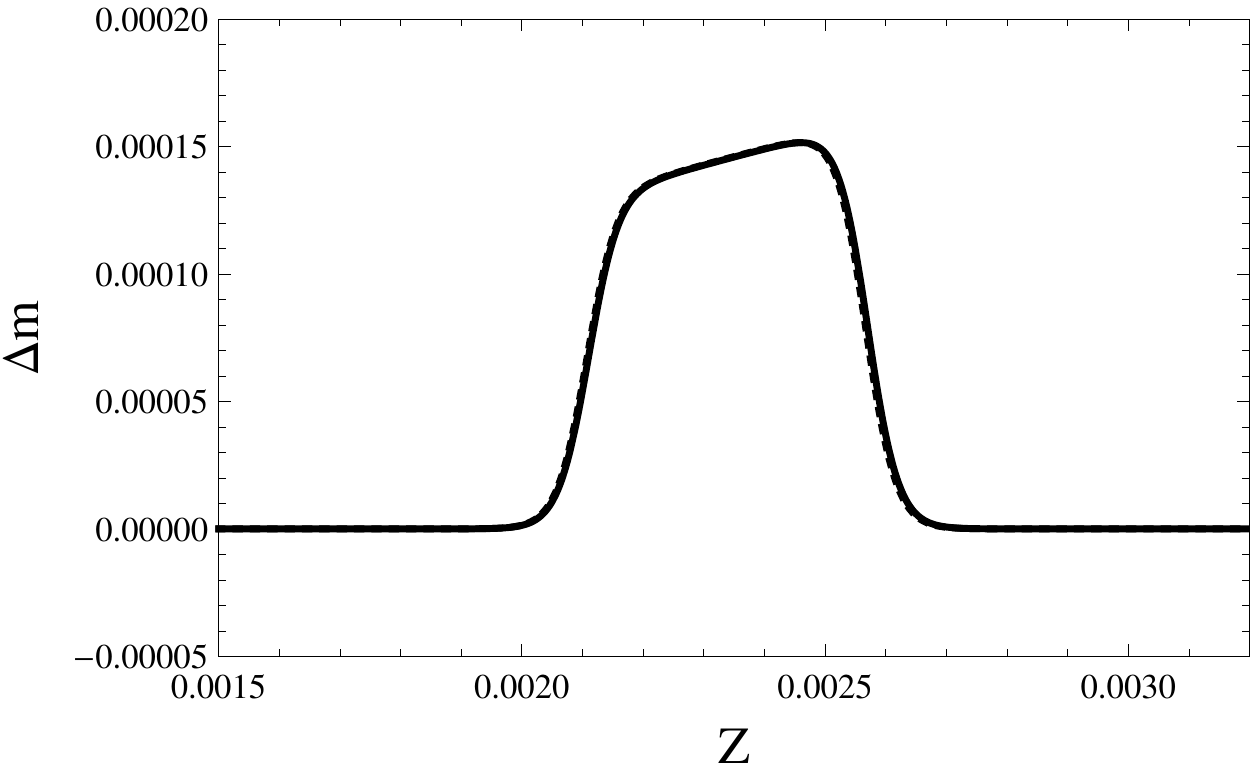}~~~~~~~~~~
 	\label{fig:Under10Mpc}
}

\caption{\label{fig:10Mpc} On the left side, the magnification is plotted for the under and over density at $d = 10 $ Mpc from the observer. On the right side, the difference in the distance modulus $\Delta m=5\,\log_{10}(d_A / \bar d_A)$ is plotted for the same cases. Solid lines refer to the quantities for the CDM model whereas dotted lines refer to the $\Lambda$CDM model.}
\end{figure}

\begin{figure}[ht!]
\centering

 \subfigure[Overdensity at d=100 Mpc]{
	\includegraphics[height=3.8cm,width=7.5cm]{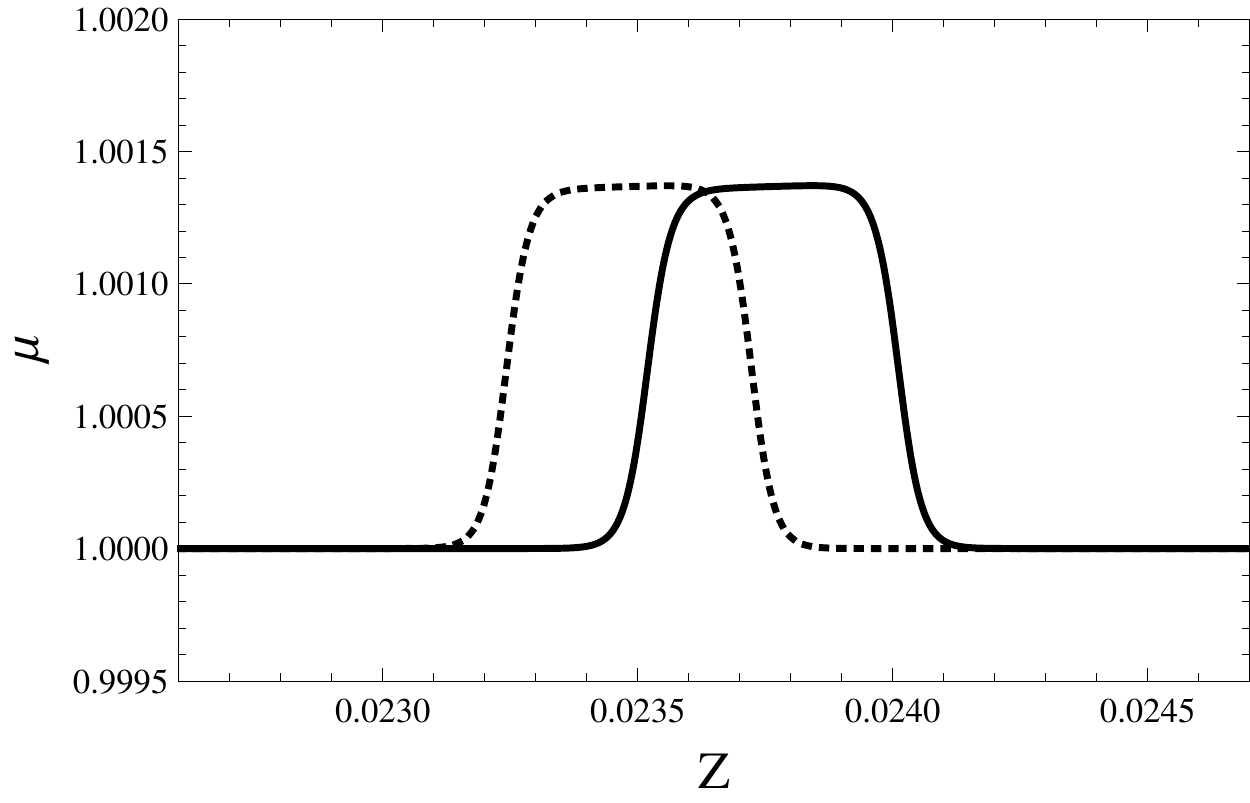}~~~~~
	\includegraphics[height=3.8cm,width=7.5cm]{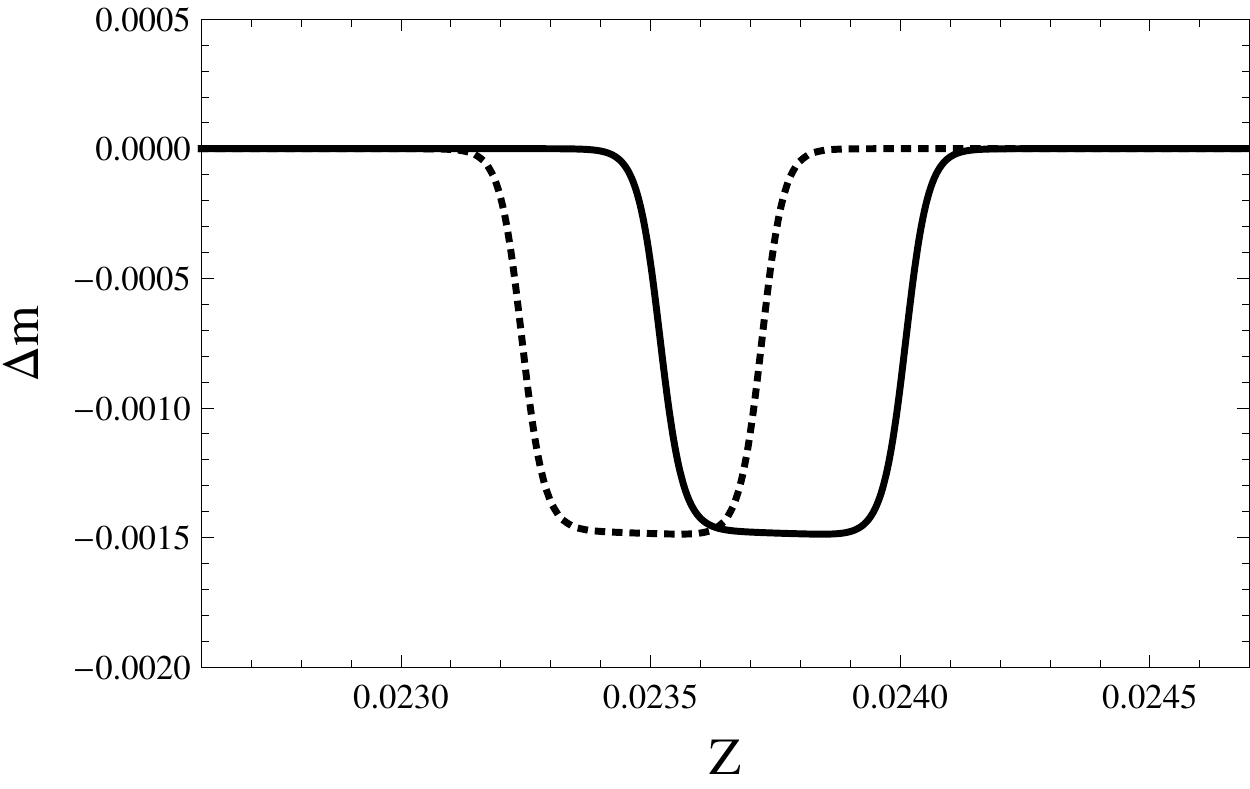}~~~~~~~~~~
	\label{fig:Over100Mpc}
 }

 \subfigure[Underdensity at d=100 Mpc]{
	\includegraphics[height=3.8cm,width=7.5cm]{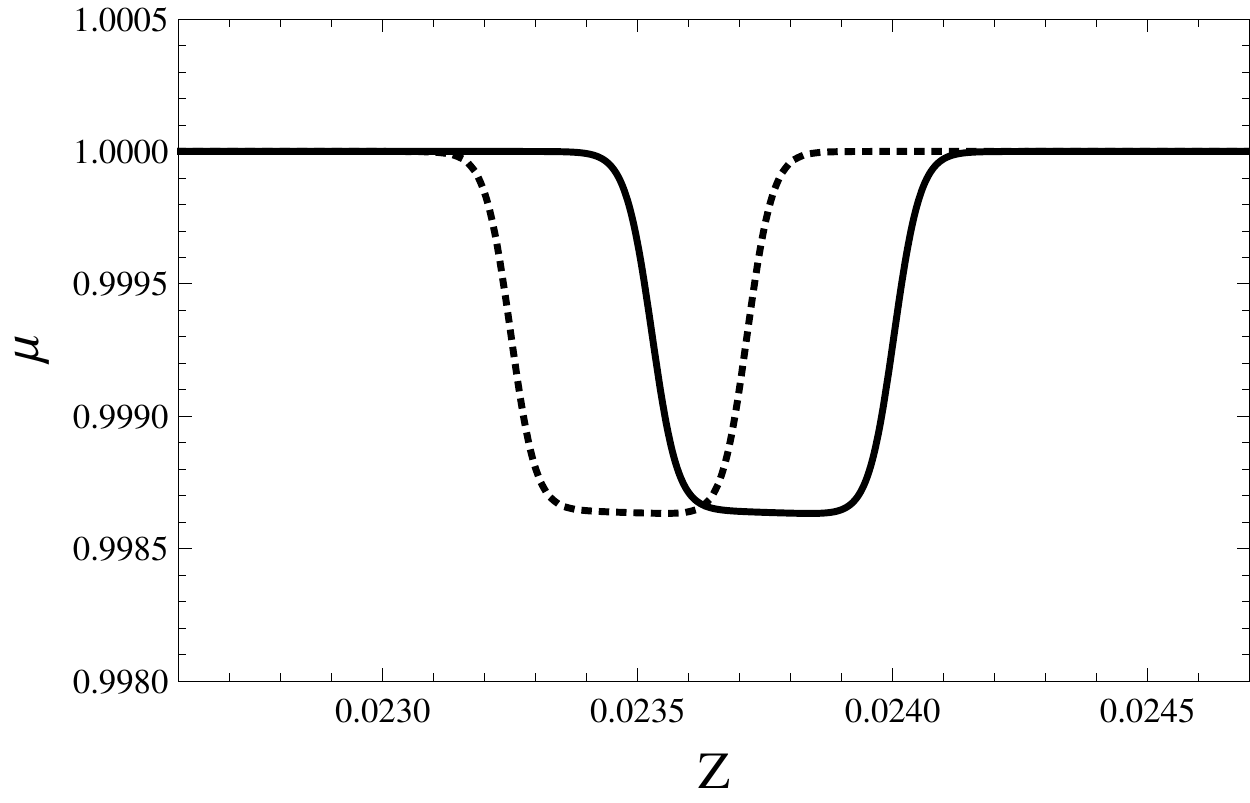}~~~~~
	\includegraphics[height=3.8cm,width=7.5cm]{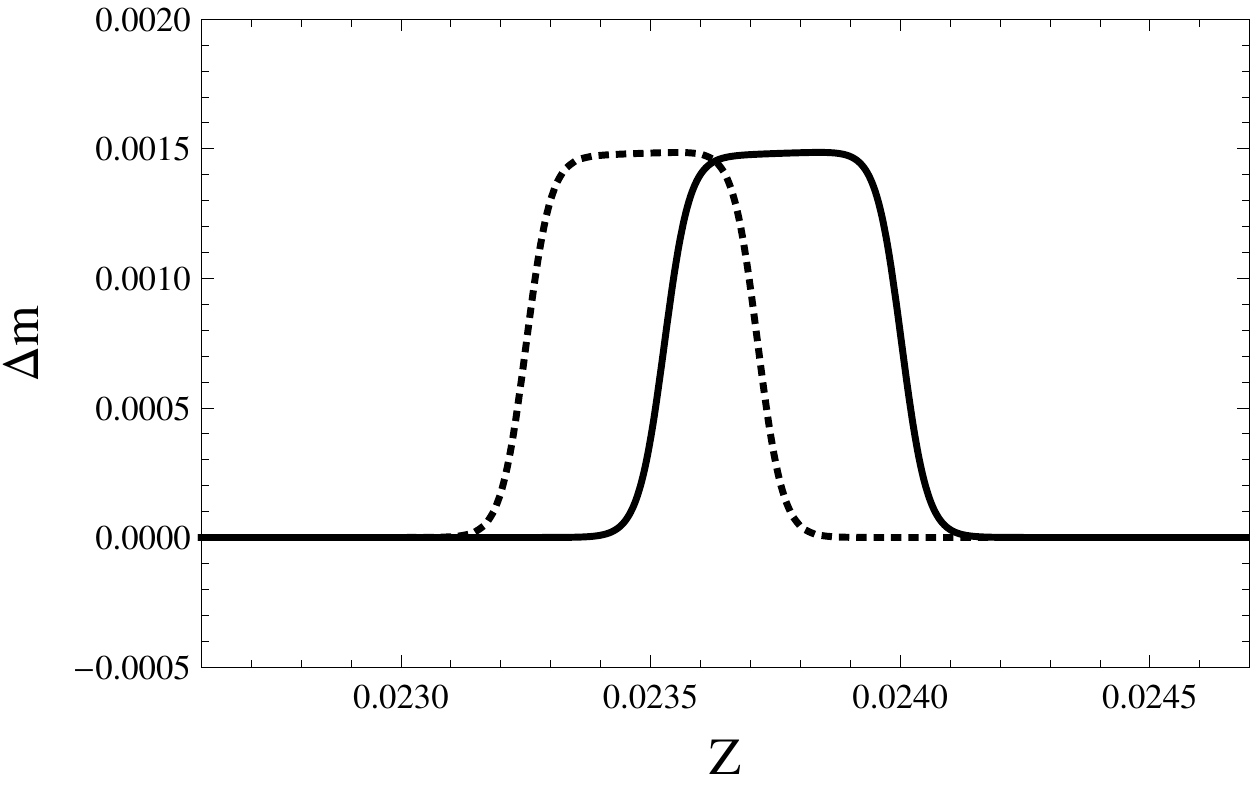}~~~~~~~~~~
	\label{fig:Under100Mpc}
 }
 
\caption{\label{fig:100Mpc} Same as Fig. \ref{fig:10Mpc} for $d = 100$ Mpc.}
\end{figure}

\begin{figure}[ht!]
\centering

 \subfigure[Overdensity at d=1000 Mpc]{
	\includegraphics[height=3.8cm,width=7.5cm]{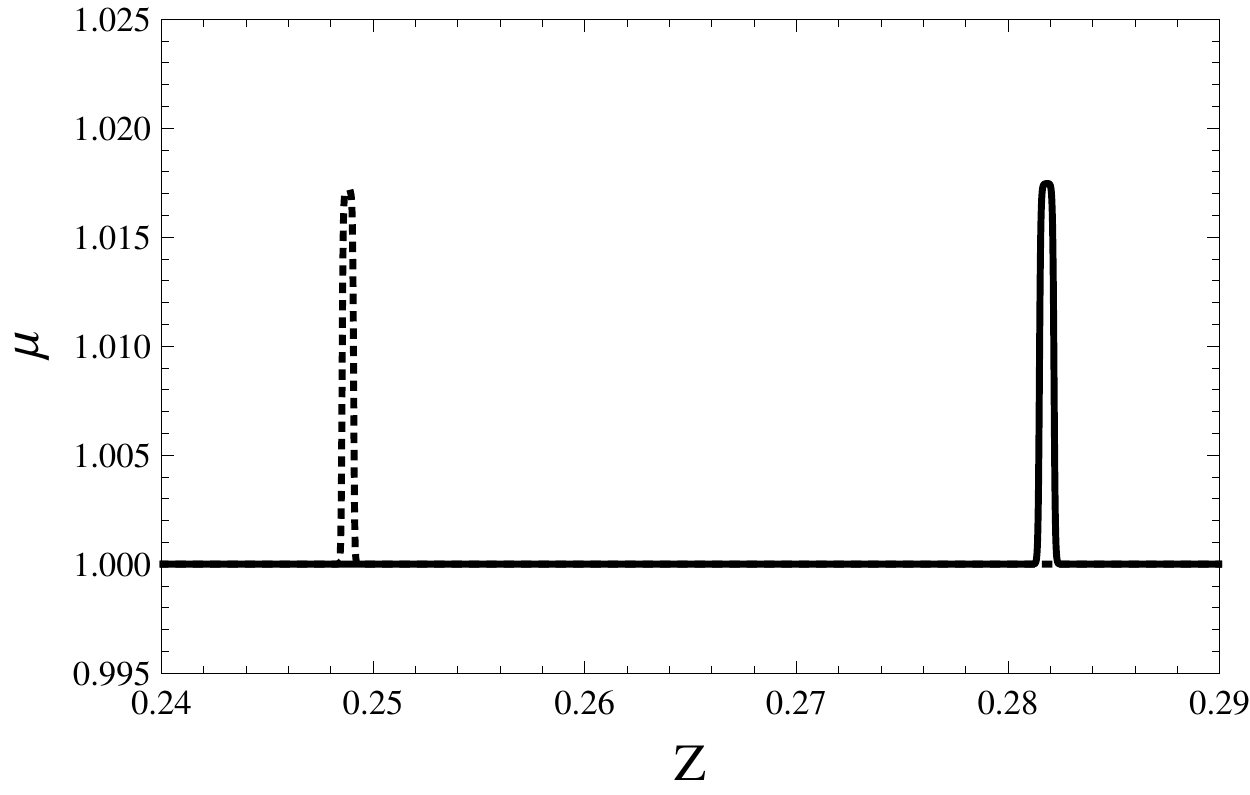}~~~~~
	\includegraphics[height=3.8cm,width=7.5cm]{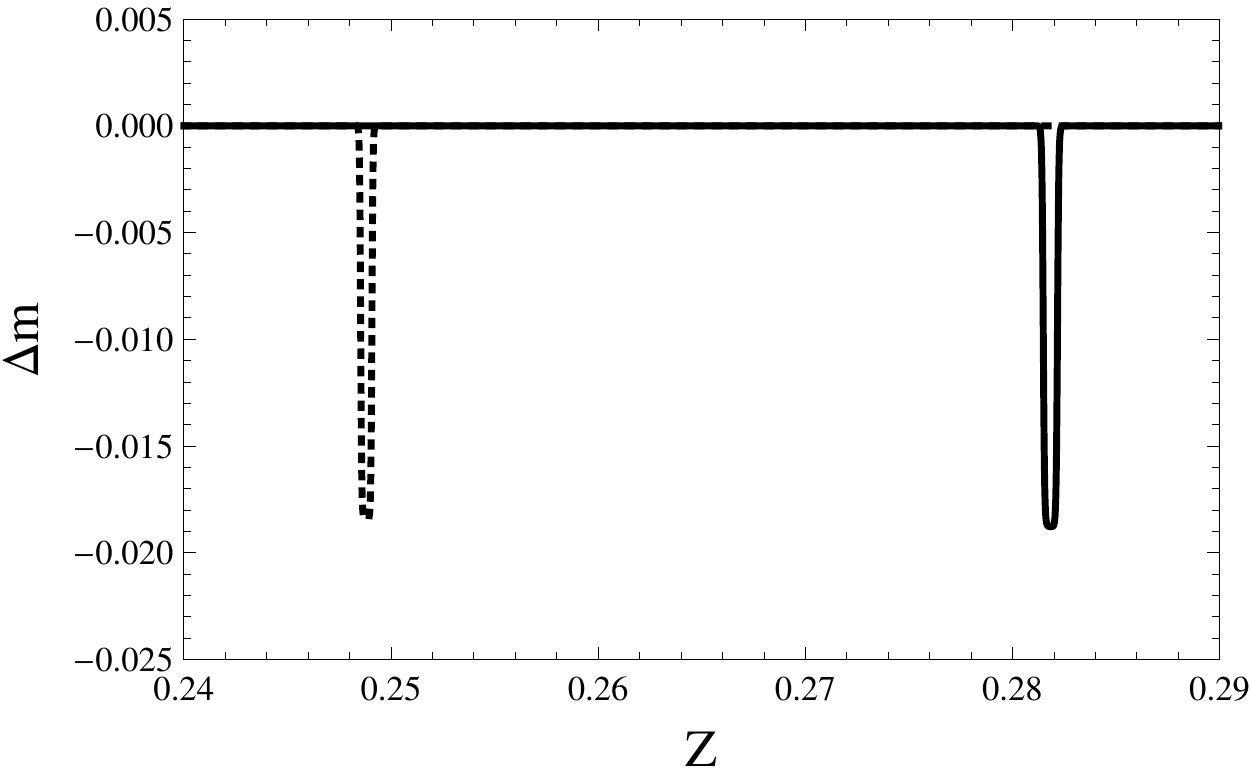}~~~~~~~~~~
	\label{fig:Over100Mpc}
 }

 \subfigure[Underdensity at d=1000 Mpc]{
	\includegraphics[height=3.8cm,width=7.5cm]{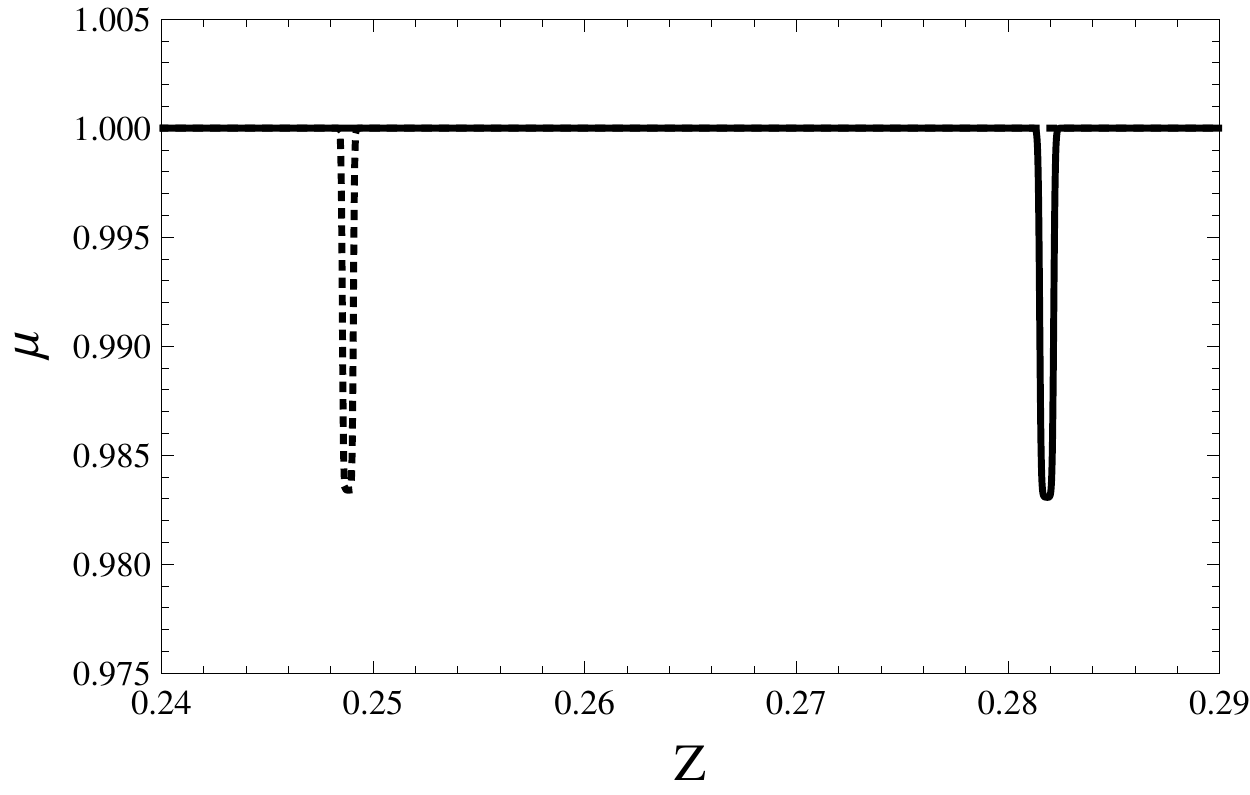}~~~~~
	\includegraphics[height=3.8cm,width=7.5cm]{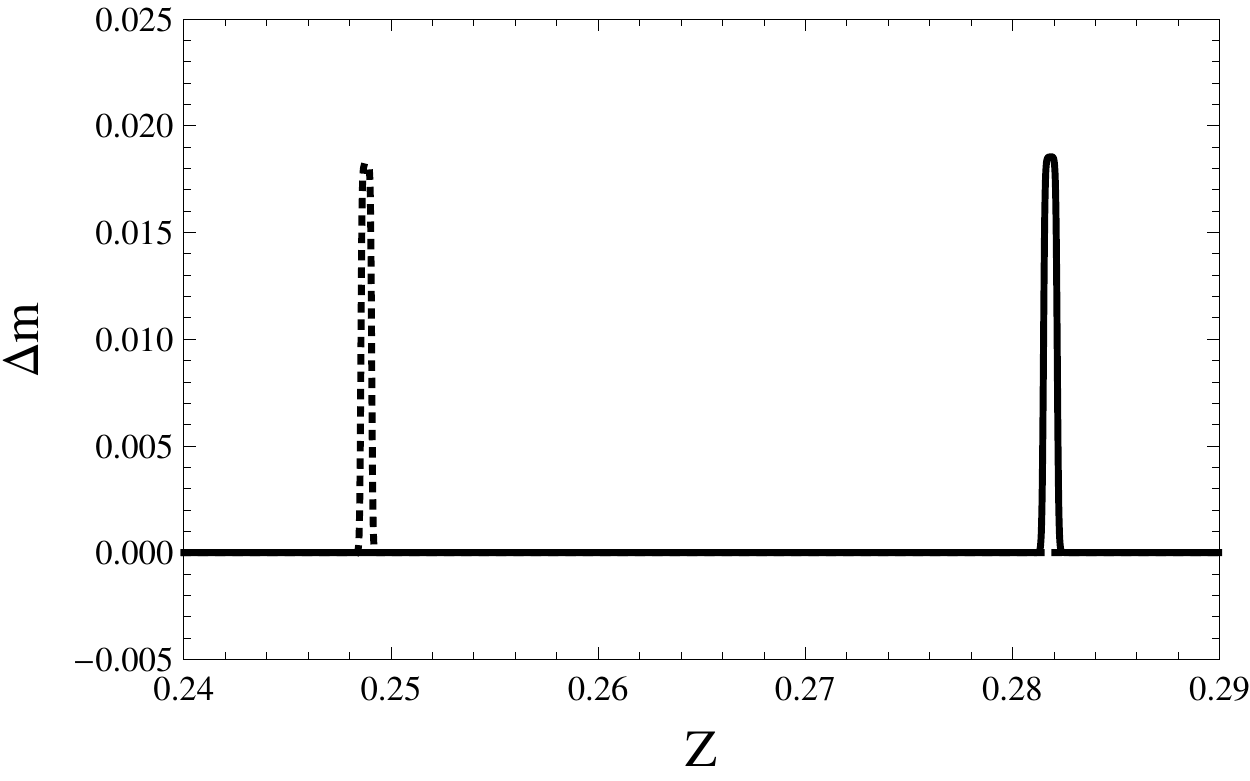}~~~~~~~~~~
	\label{fig:Under100Mpc}
 }
 
\caption{\label{fig:1000Mpc} Same as Fig. \ref{fig:10Mpc} for $d = 1000$ Mpc.}
\end{figure}

These different situations concerning the magnification $\mu$ and $\Delta m=5\,\log_{10}(d_A / \bar d_A)$ (i.e. the difference between the distance moduli of the LTB inhomogeneity case and the background one evaluated at the same redshift) have been analysed with a source first placed along the vertical axis (i.e. $\theta = \pi = \tilde \theta$) and plotted in terms of the redshift $z$. For that the geodesic equation has been numerically solved within the LTB coordinates, following \cite{Blomqvist:2009ps} (see also \cite{2011PhRvD..84d4011S,2007PhRvD..76l3004M})\,:
\beq
\frac{\di t}{\di z} = - \frac{(1+z)}{q} ~~,~~ \frac{\di r}{\di z} = \frac{p}{q} ~~,~~ \frac{\di \theta}{\di z} = \frac{J}{q A^2} ~~,~~ \frac{\di p}{\di z} = \frac{1}{q} \left[ \frac{(1-k)}{A'} \frac{J^2}{A^3} + \frac{2 \dot{A}'}{A'}p(1+z) - \left( \frac{A''}{A'} + \frac{k'}{2-2k} \right) p^2 \right] ~~,
\label{SystOfEqFromBlomqvist}
\eeq
with the constraint $q = \left[ \frac{A' \dot{A}'}{1 - k}p^2 +\frac{\dot{A} J^2}{A^3} \right]$ and $p = \di r / \di \lambda$ where $\lambda$ is our affine parameter along the geodesic (also $A' \equiv \partial_r A$, $\dot{A} \equiv \partial_t A$). $J$ is a constant angular momentum given by $J = A_0(d) \sin \tilde\theta$ and the initial conditions for the system where chosen as\,: $t = 0 ~,~ r = d ~,~ \theta = 0 ~,~ p = \cos \tilde\theta / A_0'(r)$. One should notice that this numerical integration is not necessary for our results as one can plot the lensing quantities of Eq. \rref{LensingQuantitiesExplicitInLTB} in terms of $(t,r, \theta)$. Nevertheless, the redshift is convenient as it allows us to represent our quantities in terms of only one variable. It is also a direct observable, contrary to $t$ and $r$. One should precise also that the necessity of solving the geodesic equation in LTB is due to the ``unobservable'' aspect of the LTB coordinates compared to the GLC ones. For instance, in the GLC gauge, one can easily replace the observer's proper time $\tau$ by the redshift through the relation $1+z_s = \Ups(w_0,\tau_0,\tilde\theta^a) / \Ups(w_0,\tau_s,\tilde\theta^a)$ (see \cite{P2}). This supports the idea of working instead in the GLC gauge and going to other gauges through coordinate transformations when necessary.

Our results are presented in Figs. \ref{fig:10Mpc}, \ref{fig:100Mpc} and \ref{fig:1000Mpc}. The solid lines refer to the CDM model while the dotted lines show the results for the $\Lambda$CDM case. As noticeable in the plots, the deviation from homogeneity is of the same order of magnitude in both models. It is independent from the value of $\Lambda$ (for $r_0$ small). Moreover, for $d=10$ Mpc, corrections appear at the same redshift for both models (Fig.~\ref{fig:10Mpc}). This means that the deviation from homogeneity is insensitive to the value of $\Lambda$ for small distances. However, when the inhomogeneous region is placed at $d=100$ Mpc, differences in terms of redshift appear (and even more clearly for $d=1000$ Mpc). For example, the greater the distance, the higher the correction. Indeed, at $d=10$ Mpc, the correction to the distance modulus is almost $0.015\%$ (see maximum in Fig.~\ref{fig:10Mpc}), while at $d=100$ Mpc $\Delta \mu$ reaches $\sim 0.15\%$ (Fig.~\ref{fig:100Mpc}) and for $d=1000$ Mpc the deviation from the homogeneous prediction is almost $1.5\%$. This is the direct consequence of our choice to present here a decaying mode ($k(r) = 0$). We also notice that the redshift for corrections due to large scale inhomogeneities in $\Lambda$CDM is lower than the analogous case for CDM. We interpret this by the fact that we impose the initial condition $r=d$ in the numerical resolution of Eq. \rref{SystOfEqFromBlomqvist} and this distance corresponds to a lower redshift in the $\Lambda$CDM scenario with respect to the CDM one.

\begin{figure}[ht!]
\centering

 \subfigure{
	\includegraphics[width=8cm]{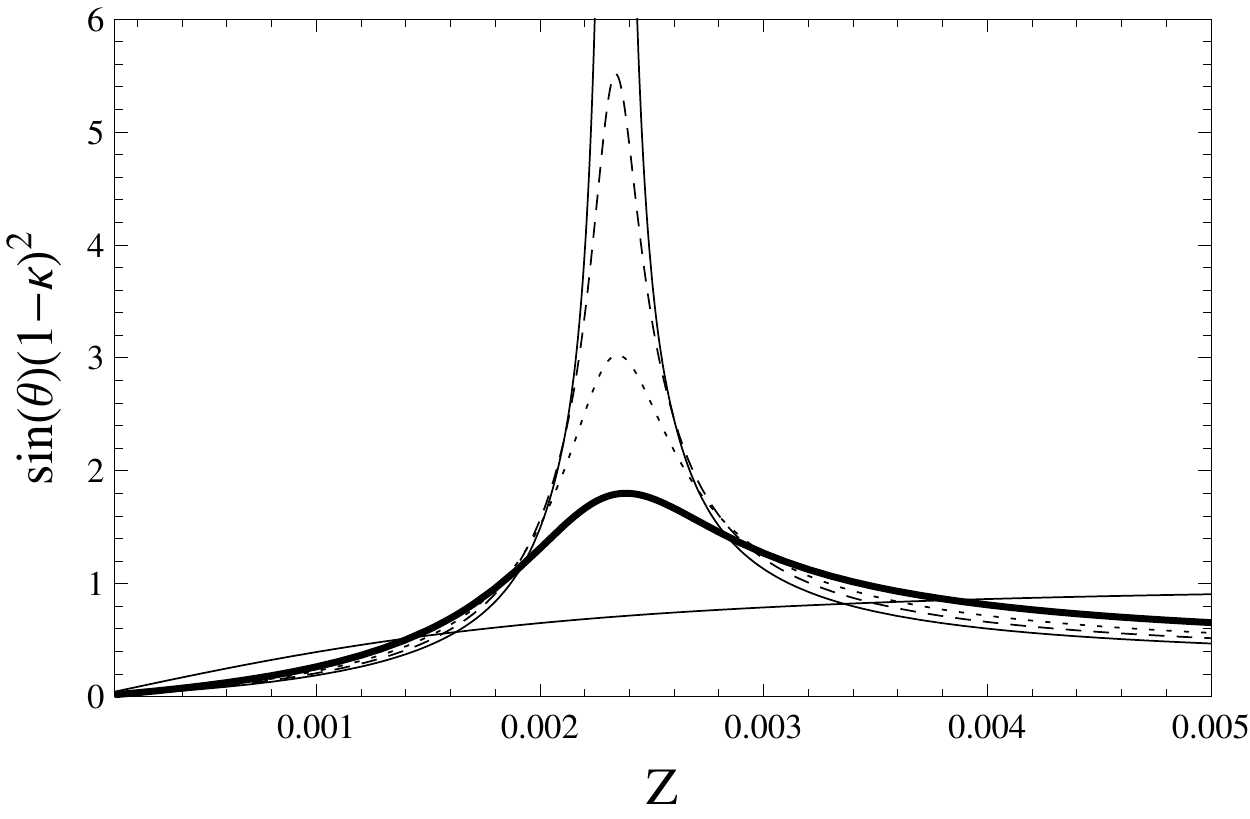}~~~~~
		\label{fig:CompConv}
	\includegraphics[width=8cm]{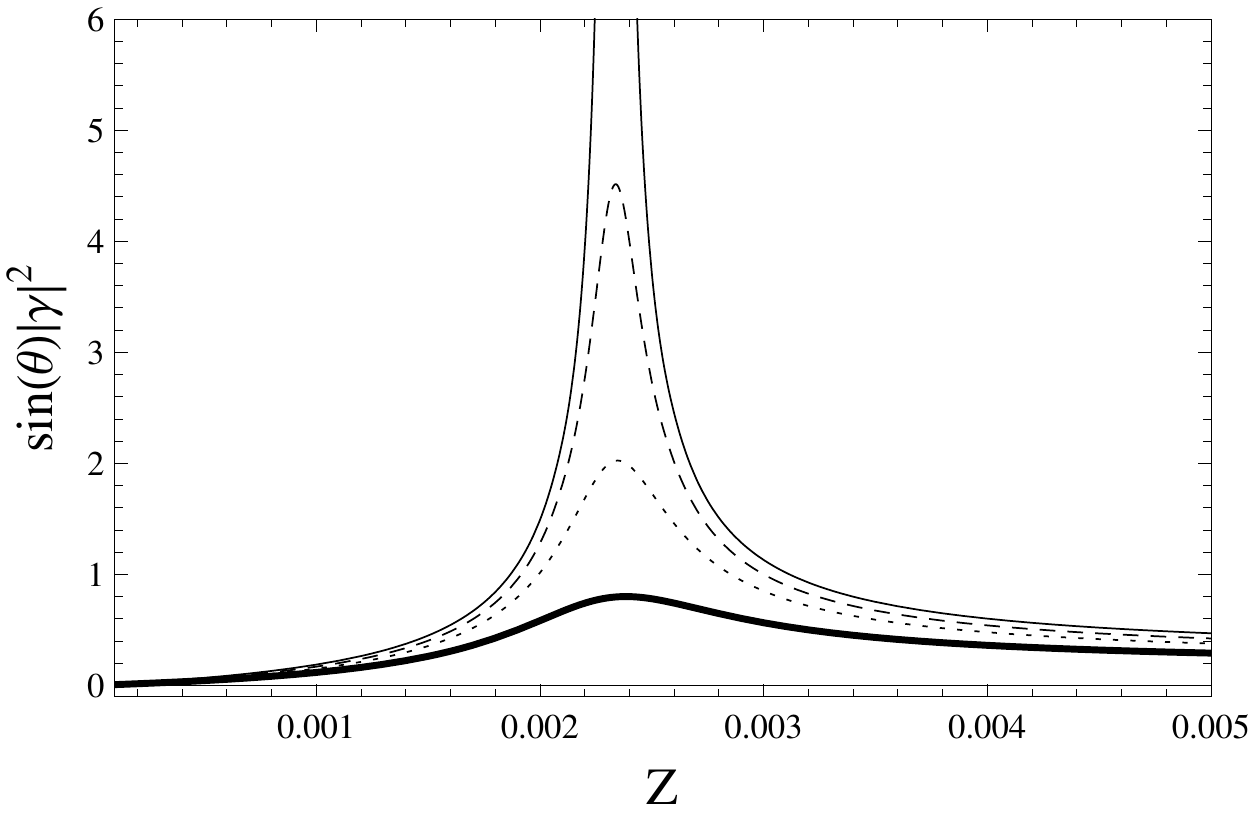}
		\label{fig:CompSh}
 }

 \subfigure{
	\includegraphics[width=8cm]{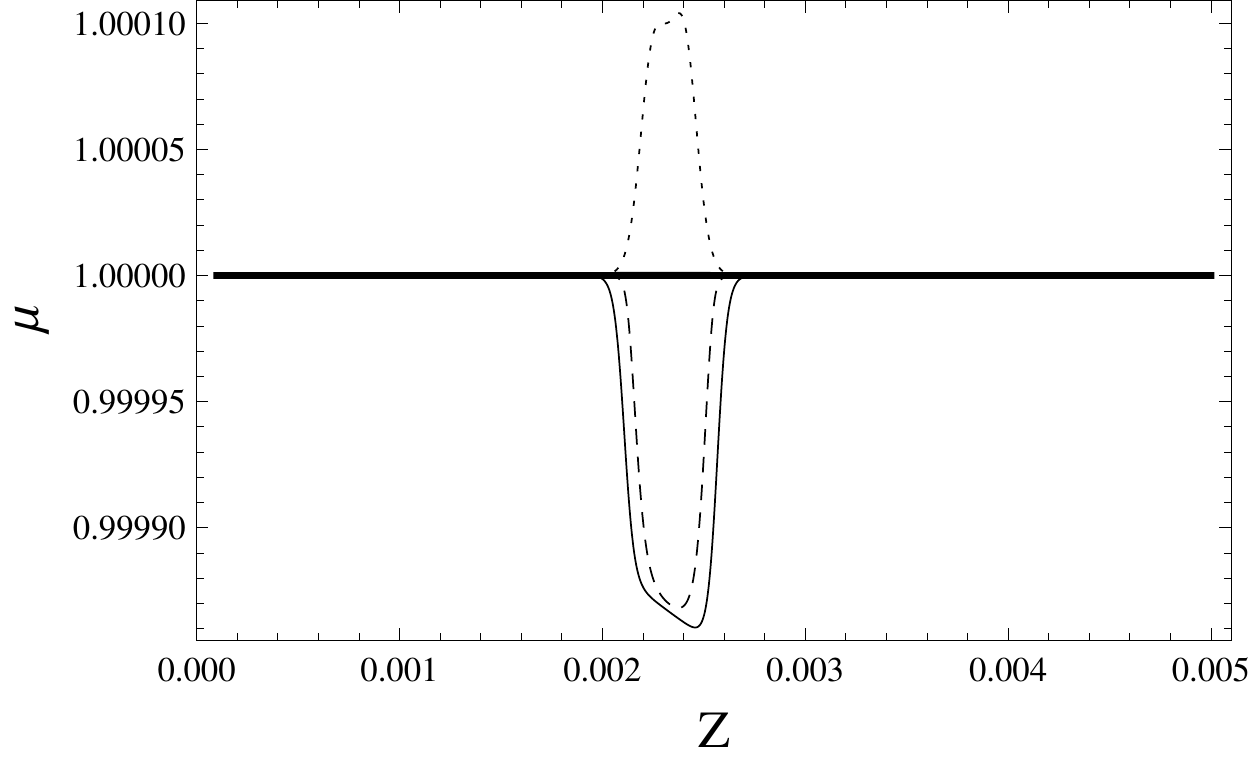}
		\label{fig:CompMagn}
 }
 
\caption{\label{fig:DifferentAngles} Redshift evolution of the magnification, convergence and shear for 5 different angles of observation\,: $\tilde \theta = $ $\pi$ and $\pi - \arcsin(10 \, r_o / d)$ (thin), $\pi - \arcsin(2 \, r_o/d)$ (thick), $\pi - \arcsin(r_o/d)$ (dotted) and $\pi - \arcsin(r_o/2 d)$ (dashed). The case represented here is the one of an underdensity with $r_o = 1 \, \rm{Mpc}$ situated at $d = 10 \, \rm{Mpc}$ and within a $\Lambda$CDM model.
\bigskip
}
\end{figure}

We finally present our results of the LTB application for a source position which is not aligned with the vertical axis (defining the direction of the observer from the center of the bubble). For that we studied the redshift evolution of $\mu$, $\sin \theta (1 - \kappa)^2$ and $\sin \theta |\hat \gamma|^2$ at different values of the angle $\tilde \theta$, namely $\tilde \theta = \pi - \arcsin(\{ 10, 2, 1, 0.5, 0 \} \times r_o/d)$. One multiplied here the values of the convergence and shear by $\sin \theta$ in order to get rid of the coordinate divergence at $\tilde{\theta} = \pi = \theta$ (see Eq. \rref{LensingQuantitiesExplicitInLTB}). These plots are obtained again after solving the geodesic equation as mentioned above and are shown in Fig. \ref{fig:DifferentAngles} (respectively in thin, thick, dotted and dashed lines). We can interpret these curves in the following way. The solid thin curves correspond to the angle far away from the bubble ($\tilde \theta = \pi - \arcsin(10 \, r_o/d)$) and at its center ($\tilde \theta = \pi$). The effect of the bubble is almost inexistent in the first case and the values are very close to $\mu =1$, $\kappa=0$ and $|\hat \gamma | =0$ (taking into account  the $\sin\theta$ term in the first two plots), as it is expected for the homogeneous case. The second case shows the maximal effect from the bubble, i.e. when the photons go though the longest part of it. In the dashed line we aim roughly at the half-radius of the bubble ($\tilde \theta = \pi - \arcsin(r_o/2d)$) and this case is quite similar to the bubble center as the profile determined by $H_0(r)$ is already reaching its central value for such an angle. These cases correspond to a demagnification ($\mu < 1$) of the source due to the underdensity and we can see in Fig. \ref{fig:DifferentAngles} that the magnification becomes $>1$ for the line of sight pointing to the border of the inhomogeneous region ($\tilde \theta = \pi - \arcsin(r_o/d)$, dotted curve), showing a magnification of sources. For a line of sight at an angle close to the angular size of the void ($\tilde \theta = \pi - \arcsin(2 \, r_o/d)$, thick line) the magnification equals one, but we can interestingly notice that the shear and the convergence are still non-zero because of their smooth evolution in terms of redshift. In this case and outside the bubble, one can show that the deviation from the homogeneous scenario is equally shared between the shear and the convergence in such a way that there is not effect on magnification. We have checked that all the curves in Fig. \ref{fig:DifferentAngles} satisfy the relation $|\hat{\gamma}|^2=(1-\kappa)^2-\mu^{-1}$.

\begin{figure}[ht!]
\centering
	\includegraphics[width=11cm]{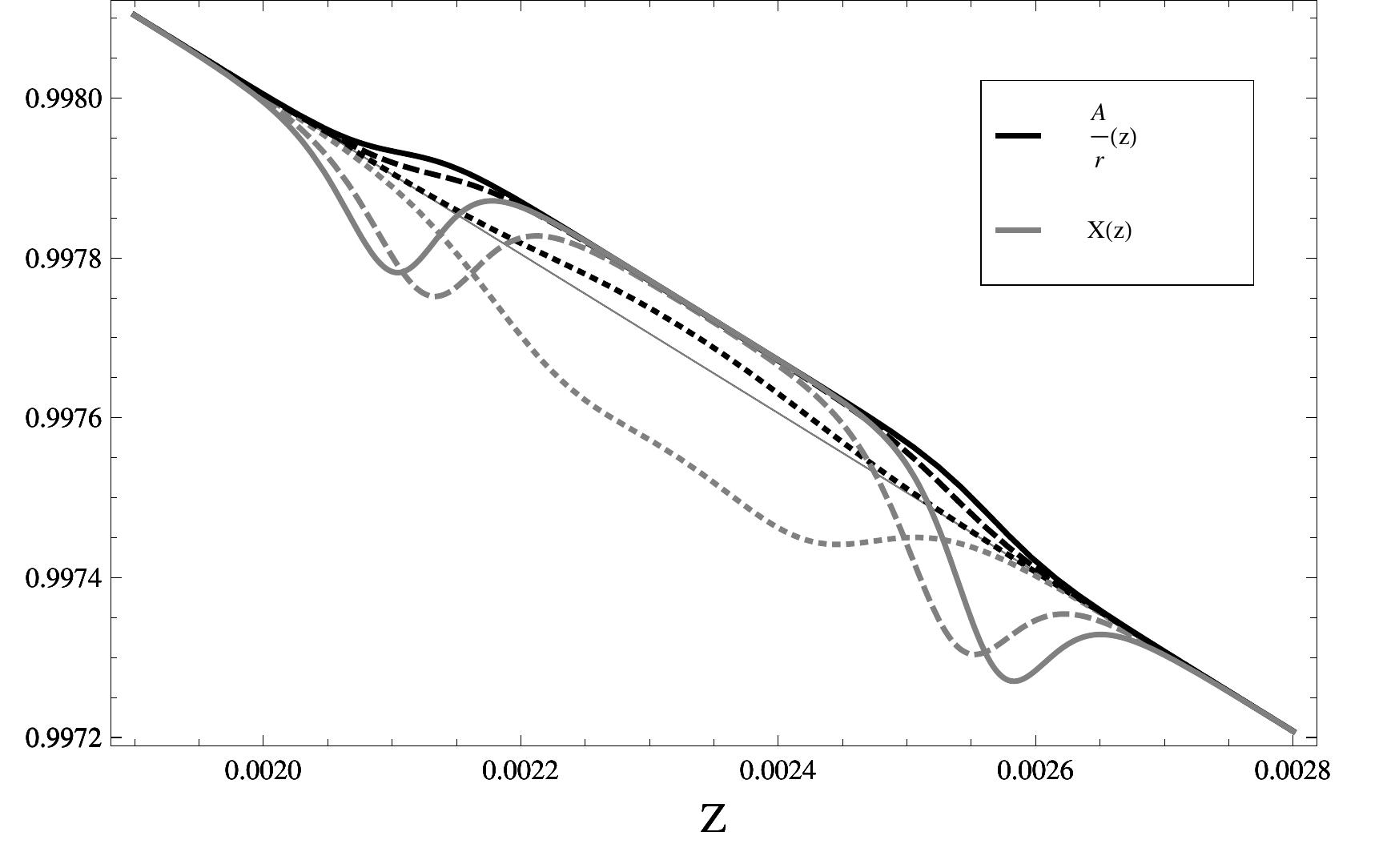}
\caption{\label{fig:NoDivergence} $A(t,r)$ (black) and $X(t,r)$ (gray) computed from our ansatz of Eq. \rref{AnsatzH0} in the case of an underdensity with $r_o = 1 \, \rm{Mpc}$ situated at $d = 10 \, \rm{Mpc}$ and within a $\Lambda$CDM model. We consider different angles of observation\,: $\tilde \theta = $ $\pi$ (thin), $\pi - \arcsin(r_o/d)$ (dotted) and $\pi - \arcsin(r_o/2 d)$ (dashed). The case $\pi - \arcsin(10 \, r_o / d)$ (thin gray) is also displayed to see the consistency with the background values.}
\end{figure}

To close and sum up this section, we have presented here the case of an uncompensated LTB under/overdensity model with $k(r) = 0$ (i.e. only the decaying mode). We intend to address the more general case $k(r) \neq 0$ in a forthcoming publication, again through the angle of the GLC coordinates. We have also checked that our LTB quantities $A(t,r)/r$ and $X(t,r)$ do not diverge, as shown for an underdensity at $d = 10 \, \rm{Mpc}$ from its observer in a $\Lambda$CDM background (same case as Fig. \ref{fig:DifferentAngles}). The results are presented in Fig. \ref{fig:NoDivergence} and indicate that our metric functions are free from real singularities (as opposed to a coordinate singularities).

We can conclude this section by noticing that the plots presented here are the consequence of the exact and explicit derivation of lensing quantities given in Sec. \ref{Sec4}. We recall that the resolution of the geodesic deviation equation is necessary only to show the lensing quantities in terms of the redshift, a complication related to the LTB coordinates. Up to our knowledge, our approach is the first to attempt such an explicit resolution for an off-center observer in an LTB model and that is possible thanks to the great flexibility of the geodesic light-cone coordinates. This approach can be generalized to other kinds of voids and other types of coordinates and it would be interesting to consider the cosmological applications of these derivations of lensing quantities.

\newpage

\section{Conclusions and Outlook}
\label{Sec6}
\label{SecConcl}

We have presented in this paper the explicit expressions of several lensing quantities contained in the amplification matrix (magnification, convergence and shear) and the deformation matrix (optical scalars). These results were obtained by the use of the Jacobi map expressed first through the zweibeins of the Sachs basis and then within the geodesic light-cone coordinates. We have shown that some particular combinations of the lensing quantities take an elegant expression in the GLC gauge and the explicit GLC-form of the zweibeins is not needed (though it was given for completeness in Appendix \ref{SecAppendixA}). These expressions are general and can be applied to any inhomogeneous model of Universe as long as no caustic forms on the past light-cone of the observer. We have also seen through these computations the interest of the GLC gauge in dealing with light propagation, emphasizing that the geodesic equation being trivial in this set of coordinates also simplifies the handling of lensing variables.

Let us comment briefly on the presence of caustics. In this situation one sees the creation of critical points or lines on a given sphere $\Sigma(w,\tau)$ embedded in the past light-cone. These singularities appear when $\det J^{A}_{B}$ is equal to zero. In such a case the Jacobi map is not an invertible matrix and complicated situations with multiple images and infinite magnification can be observed. We can also see (from Eqs. \rref{dAJAB}, \rref{JABandCaB} and \rref{dAInGLC}) that this happens when $\gamma \equiv \det \gamma_{ab} = 0$ in the GLC gauge and we then have a case of strong lensing. The estimation of the angular distance is not anymore valid on these critical lines \cite{Ellis:1998ha} and a more involved system of coordinates is necessary. The generalisation of the GLC gauge to that case appears as a hard task, despite its simplifying properties in the case of weak lensing.

In the second part of this paper we have applied our general formulas to the well-known case of a LTB metric with an off-center observer. These relations, especially the lensing quantities, were obtained directly after expressing the non-perturbative transformation between LTB and GLC coordinates. They are general and to our knowledge the most explicit expressions presented up to now and they could be used for a broad range of applications, in particular for the exact treatment of light propagation in inhomogeneous universes. We then restricted our study to an LTB over/underdensity (or ``bubble'') entirely determined by its local Hubble factor within CDM and $\Lambda$CDM backgrounds. We presented the variations of the magnification $\mu$ and the distance modulus difference $\Delta m$ for an over/underdense bubble, pointing at the center of it. We found the correction induced on the magnification by the bubble to be very small at short distances ($\sim 0.015 \%$ for $d=10$ Mpc) at more significant at large distances ($\sim 1.5 \%$ for $d=1000$ Mpc) in the $\Lambda$CDM case as well as CDM (the difference being negligible for such small under/overdensities). We also showed the consistency of our results by presenting the evolution of the quantities, including the shear and the convergence, when we point at different angles from the center of the bubble. We finally checked that our quantities are well-behaved and that no divergence appears.

To finish, this work improves our understanding of lensing theory and presents some of its most important quantities in an elegant framework. It also illustrates -- once again -- the general usefulness of the GLC gauge for deriving theoretical expressions as well as working out with more applied situations. We illustrated our results through the LTB model but we can imagine other applications such as, for example, the computation of the lensing quantities in an anisotropic Bianchi model or at second order in perturbations in the Poisson gauge. Lensing benefits from a growing interest in cosmology today, in experimental (e.g. \cite{Ade:2013tyw,2012ApJ...756..142Vcut,Das:2013zf,Amendola:2012ys, Abbott:2005bi, 2009arXiv0912.0201L}) as well as computational aspects (e.g. \cite{Metcalf:2013kya, Petkova:2013yea}), these expressions and their wide applicability for different models could hence give a nice framework for new lensing predictions and simulations.

\vspace{2cm}

\section*{ACKNOWLEDGMENTS}
The authors of this paper wish to thank G. Veneziano, B. Metcalf, G. Marozzi, M. Gasperini, P. Fleury and D. Leier for helpful discussions. FN's research is supported by the project GLENCO, funded under the FP7, Ideas, Grant Agreement n. 259349. GF's work is supported by the research grant ``Theoretical Astroparticle Physics'' number 2012CPPYP7 under the program PRIN 2012 funded by the Ministero dell'Istruzione, Universit\`{a} e della Ricerca (MIUR) and by the Italian Istituto Nazionale di Fisica Nucleare (INFN) through the ``Theoretical Astroparticle Physics'' project. GF also wish to thank the University of Geneva for its hospitality during part of the elaboration of this document.

\vspace{2cm}

\newpage

\appendix
\begin{appendices}

\section{Explicit solution for the zweibeins}
\label{SecAppendixA}

The general expression of the Sachs basis is given by the resolution of Eqs. \rref{eq:Sachs}, or equivalently in the GLC gauge by the conditions $\gamma_{ab} s_A^a s_B^b = \delta_{AB}$ and $\nabla_\lambda s^a_A = 0$ (see Eq. \rref{ParallelTransport}). In this resolution, the first condition allows us to fix three components of the zweibeins in terms of the remaining one and obtain\,:
\bea
s^a_1=\left( \alpha,\frac{-\alpha\gamma_{12}+\sqrt{\gamma_{22}-\alpha^2\gamma}}{\gamma_{22}} \right)\qquad,\qquad s^a_2=\left( -\sqrt{\frac{\gamma_{22}-\alpha^2\gamma}{\gamma}},\frac{\alpha\gamma+\gamma_{12}\sqrt{\gamma_{22}-\alpha^2\gamma}}{\gamma_{22}\sqrt{\gamma}} \right) ~~.
\eea
The component $\alpha$ is fixed by the second condition, i.e. the parallel transport condition, which can also be written as $\epsilon^{AB}\dot s^a_A s_{aB}=0$ and which translates as\,:
\beq
\label{eq:parallel}
\dot\alpha=-\frac{\sqrt{\gamma_{22}-\alpha^2\gamma}\left( \gamma_{22}\dot\gamma_{12}-\gamma_{12}\dot\gamma_{22} \right)+\alpha\left( (\gamma_{22})^2\dot\gamma_{11}-2 \gamma_{12}\gamma_{22}\dot\gamma_{12}+(\gamma_{12})^2\dot\gamma_{22} \right)}{2\gamma_{22}\gamma} ~~.
\eeq
This last equation can be rewritten in terms of $\beta \equiv \beta(\tau,w,\tilde\theta^a)$ such that $\alpha=\cos\beta\sqrt{\frac{\gamma_{22}}{\gamma}}$. It then becomes\,:
\beq
\dot\beta=\frac{\gamma_{22}\dot\gamma_{12}-\gamma_{12}\dot\gamma_{22}}{2\,\gamma_{22}\sqrt{\gamma}}=\sqrt{\gamma}\frac{\gamma^{1c}\dot\gamma_{c2}}{2\,\gamma_{22}}
\eeq
and its solution is simply given by\,:
\beq
\label{eq:parallelAngle}
\beta=\int\sqrt{\gamma}\frac{\gamma^{1c}\dot\gamma_{c2}}{2\,\gamma_{22}}\di \tau
\eeq
where we have chosen the integration constant to be equal to 0. This allows us to say that $\beta=0$ for a diagonal $\gamma_{ab}$ (as it is the case for our LTB application of Sec \ref{Sec4}), i.e. $s^a_1=\left( \gamma_{11}^{-1/2},0 \right)$ and $s^a_2=\left( 0,\gamma_{22}^{-1/2} \right)$. Therefore, the general expression for the zweibeins is\,:
\beq
\label{eq:explicitZweibein}
s^a_1=\left( \cos\beta \sqrt{\frac{\gamma_{22}}{\gamma}}, -\cos\beta\frac{\gamma_{12}}{\sqrt{\gamma\gamma_{22}}}+\frac{\sin\beta}{\sqrt{\gamma_{22}}} \right)\qquad,\qquad s^a_2=\left( -\sin\beta \sqrt{\frac{\gamma_{22}}{\gamma}},\frac{\cos\beta}{\sqrt{\gamma_{22}}}+\sin\beta\frac{\gamma_{12}}{\sqrt{\gamma\gamma_{22}}} \right) ~~.
\eeq

As shown in Appendix A of \cite{Fanizza:2013doa}, Eq. \eqref{eq:parallelAngle} gives the right angle in order to satisfy the parallel transport condition for $s^a_A$. In fact, Eq. \eqref{eq:explicitZweibein} can also be viewed as a rotation of an angle $\beta$ on the simplest solution of the condition $\gamma_{ab}s^a_As^b_B=\delta_{AB}$, namely the solution given by Eq. \eqref{eq:explicitZweibein} with $\beta = 0$. Indeed, after some simple algebraic manipulations, it can be shown that $s^a_A=\mathscr{R}^B_A \, \tilde s^a_B$, where
\beq
\tilde s^a_1=\left( \sqrt{\frac{\gamma_{22}}{\gamma}}, -\frac{\gamma_{12}}{\sqrt{\gamma\gamma_{22}}} \right) ~~~~~,~~~~~ \tilde s^a_2=\left( 0 \, , \frac{1}{\sqrt{\gamma_{22}}} \right) ~~~~~,~~~~~
\mathscr{R}^B_A=\left(
\begin{array}{cc}
\cos\beta&\sin\beta\\
-\sin\beta&\cos\beta
\end{array}
\right)
\eeq
and $\mathscr{R}^B_A$ corresponds to the real irreducible representation of the symmetry group $U(1)$ on the zweibeins.

\section{Complement on the explicit expressions of lensing quantities}
\label{SecAppendixB}

The expression of the Jacobi map in Eq. \rref{JABandCaB} with the relations presented in Eq. \rref{ExpressionsLensingQuantities} lead to the convergence, vorticity and shear expressed in the GLC gauge and in terms of the angle $\beta$ described in Appendix \ref{SecAppendixA}. We find\,:
\bea
\label{eq:amplificationQuantities1}
\kappa &=& 1-\frac{u_{\tau_o}}{\bar d_A\sqrt{\gamma_{22}\gamma_{22o}}\left( \det^{ab}\dot\gamma_{ab} \right)_o}\left\{ \cos\left( \beta-\beta_o \right)\left[  \gamma_{22}\gamma_{22o}\,\dot\gamma_{11o}+\left( \gamma_{12}\gamma_{12o}+\sqrt{\gamma\,\gamma_o} \right)\,\dot\gamma_{22o}-\left( \gamma_{12o}\gamma_{22}+\gamma_{12}\gamma_{22o} \right)\dot\gamma_{12o} \right]\right.\nonumber\\
&-&\left.\sin\left( \beta-\beta_o \right)\left[ \left(\gamma_{22}\sqrt{\gamma_o}-\gamma_{22o}\sqrt{\gamma}\right)\dot\gamma_{12o}+\left( \gamma_{12o}\sqrt{\gamma}-\gamma_{12}\sqrt{\gamma_o} \right)\dot\gamma_{22o} \right]\right\} ~~, \nonumber\\
\hat{\omega}
&=&\frac{u_{\tau_o}}{\bar d_A\sqrt{\gamma_{22}\gamma_{22o}}\left(\det^{ab}\dot\gamma_{ab}\right)_o}\left\{ \sin\left( \beta-\beta_o \right)\left[  \gamma_{22}\gamma_{22o}\,\dot\gamma_{11o}+\left( \gamma_{12}\gamma_{12o}+\sqrt{\gamma\,\gamma_o} \right)\,\dot\gamma_{22o}-\left( \gamma_{12o}\gamma_{22}+\gamma_{12}\gamma_{22o} \right)\dot\gamma_{12o} \right]\right.\nonumber\\
&+&\left.\cos\left( \beta-\beta_o \right)\left[ \left(\gamma_{22}\sqrt{\gamma_o}-\gamma_{22o}\sqrt{\gamma}\right)\dot\gamma_{12o}+\left( \gamma_{12o}\sqrt{\gamma}-\gamma_{12}\sqrt{\gamma_o} \right)\dot\gamma_{22o} \right]\right\} ~~, \nonumber\\
\eea
\bea
\label{eq:amplificationQuantities2}
\hat{\gamma}_1
&=&\frac{u_{\tau_o}}{\bar d_A\sqrt{\gamma_{22}\gamma_{22o}}\left(\det^{ab}\dot\gamma_{ab}\right)_o}\left\{ \cos\left( \beta+\beta_o \right)\left[ \gamma_{22}\gamma_{22o}\dot\gamma_{11o}+\left( \gamma_{12}\gamma_{12o}-\sqrt{\gamma\gamma_o} \right)\dot\gamma_{22o}-\left( \gamma_{22}\gamma_{12o}+\gamma_{12}\gamma_{22o} \right)\dot\gamma_{12o} \right]\right.\nonumber\\
&+&\left.\sin\left( \beta+\beta_o \right)\left[ \left( \gamma_{22o}\sqrt{\gamma}+\gamma_{22}\sqrt{\gamma_o} \right)\dot\gamma_{12o}-\left( \gamma_{12o}\sqrt{\gamma}+\gamma_{12}\sqrt{\gamma_o} \right)\dot\gamma_{22o} \right] \right\} ~~, \nonumber\\
\hat{\gamma}_2
&=&\frac{u_{\tau_o}}{\bar d_A\sqrt{\gamma_{22}\gamma_{22o}}\left(\det^{ab}\dot\gamma_{ab}\right)_o}\left\{ -\sin\left( \beta+\beta_o \right)\left[ \gamma_{22}\gamma_{22o}\dot\gamma_{11o}+\left( \gamma_{12}\gamma_{12o}-\sqrt{\gamma\gamma_o} \right)\dot\gamma_{22o}-\left( \gamma_{22}\gamma_{12o}+\gamma_{12}\gamma_{22o} \right)\dot\gamma_{12o} \right]\right.\nonumber\\
&+&\left.\cos\left( \beta+\beta_o \right)\left[ \left( \gamma_{22o}\sqrt{\gamma}+\gamma_{22}\sqrt{\gamma_o} \right)\dot\gamma_{12o}-\left( \gamma_{12o}\sqrt{\gamma}+\gamma_{12}\sqrt{\gamma_o} \right)\dot\gamma_{22o} \right] \right\} ~~,
\eea
and we can check that these expressions give back the results presented in Eq. \rref{LensingCombinationsInGLC}.

\bigskip

By taking the angles $\beta = \beta_o = 0$, which happens only in the singular case where $\gamma_{22}\dot\gamma_{12}-\gamma_{12}\dot\gamma_{22} = 0$, we get the simple expressions\,:
\bea
\label{eq:amplificationQuantitiesSimplified}
\kappa &=& 1-\frac{u_{\tau_o}}{\bar d_A\,\sqrt{\gamma_{22}\,\gamma_{22o}}\,(\det\dot\gamma_{ab})_o}\left[ \gamma_{22}\gamma_{22o}\,\dot\gamma_{11o}+\left( \gamma_{12}\gamma_{12o}+\sqrt{\gamma\,\gamma_o} \right)\,\dot\gamma_{22o}-\left( \gamma_{12o}\gamma_{22}+\gamma_{12}\gamma_{22o} \right)\dot\gamma_{12o} \right] ~~, \nonumber\\
\hat{\omega} &=&\frac{u_{\tau_o}}{\bar d_A\sqrt{\gamma_{22}\gamma_{22o}}\left(\det\dot\gamma_{ab}\right)_o}\left[ \left(\gamma_{12o}\sqrt{\gamma}-\gamma_{12}\sqrt{\gamma_o}\right)\dot\gamma_{22o}-\left( \gamma_{22o}\sqrt{\gamma}-\gamma_{22}\sqrt{\gamma_o} \right)\dot\gamma_{12o} \right] ~~, \nonumber\\
\hat{\gamma}_1 &=& \frac{u_{\tau_o}}{\bar d_A\,\sqrt{\gamma_{22}\,\gamma_{22o}}\,(\det\dot\gamma_{ab})_o}\left[ \gamma_{22}\gamma_{22o}\,\dot\gamma_{11o}+\left( \gamma_{12}\gamma_{12o}-\sqrt{\gamma\,\gamma_o} \right)\,\dot\gamma_{22o}-\left( \gamma_{12o}\gamma_{22}+\gamma_{12}\gamma_{22o} \right)\dot\gamma_{12o} \right] ~~, \nonumber\\
\hat\gamma_2 &=&\frac{u_{\tau_o}}{\bar d_A\sqrt{\gamma_{22}\gamma_{22o}}\left(\det\dot\gamma_{ab}\right)_o}\left[ -\left(\gamma_{12o}\sqrt{\gamma}+\gamma_{12}\sqrt{\gamma_o}\right)\dot\gamma_{22o}+\left( \gamma_{22o}\sqrt{\gamma}+\gamma_{22}\sqrt{\gamma_o} \right)\dot\gamma_{12o} \right] ~~.
\eea
and under the stronger assumption of $\gamma_{ab}$ being diagonal we get back the results presented in Eq. \rref{LensingQuantitiesInLTB} with the identity $\hat{\omega}=0$.

Similarly, the Ricci and Weyl focusing of Eq. \rref{Phi00} are recast, in the case $\beta = \beta_o = 0$, into\,:
\bea
\label{Phi00explicit}
\Phi_{00}&=& \frac{\omega^2}{4\Ups^2} \left\{ \frac{2 \Ups}{\sqrt{\gamma}} \left( \frac{(\sqrt{\gamma})^{\tdev}}{\Ups} \right)^{\tdev} - \frac{\det \dot{\gamma}_{ab}}{\gamma} \right\} ~~, \nonumber \\
\label{RePsi0explicit}
\text{Re}\Psi_0 &=& \Phi_{00} + \frac{\omega^2}{4 \Ups^2} \Bigg[ \left( \frac{\dot{\gamma}}{\gamma} + 2 \frac{\dot{\Ups}}{\Ups} \right) \frac{\dot{\gamma}_{22}}{\gamma_{22}} 
 - 2 \frac{\ddot{\gamma_{22}}}{\gamma_{22}} - \frac{\det \dot{\gamma}_{ab}}{\gamma} \Bigg] ~~, \nonumber \\
 \label{ImPsi0explicit}
\text{Im}\Psi_0 &=& \frac{\omega^2}{4 \Ups^2 \sqrt{\gamma}} \left[ \left( \frac{\dot{\gamma}}{\gamma} + 2 \frac{\dot{\Ups}}{\Ups} \right) \left( \frac{\gamma_{12} \dot{\gamma}_{22} - \gamma_{22}\dot{\gamma}_{12}}{\gamma_{22}} \right) - 2 \left( \frac{\gamma_{12} \ddot{\gamma}_{22} - \gamma_{22}\ddot{\gamma}_{12}}{\gamma_{22}} \right) \right]  ~~,
\eea
as it can also be checked from the use of Eqs. \rref{eq:SachsEq1}, \rref{eq:SachsEq2} or differently from definitions of Eq. \rref{eq:focusing}. In the simple case of a diagonal $\gamma_{ab}$ we find that $\text{Im}\Psi_0 = 0$.

\section{Limits of lensing quantities in the homogeneous FLRW case.}
\label{SecAppendixC}

The expressions of the GLC coordinates and metric elements are obtained by the comparison of Eq. \rref{GLCmetric} with the spatially flat FLRW geometry metric written in spherical coordinates $(\chi, \theta, \phi)$\,:
\beq
\label{ds2FLRW}
\di s^2_{\rm FLRW} = -\di t^2 + a^2(t) \left[\di \chi^2 + f_K^2(\chi) \left( \di \theta^2 + \sin^2 \theta \, \di \phi^2 \right) \right] ~~,
\eeq
with scale factor $a(t)$ and cosmic time $t$. These expressions are\,:
\bea
\label{FR}
&
\tau=t ~~~\mbox{(exact)}~~ ~~~~~,~~~~~ w= \chi + \eta ~~~~~,~~~~~ \tilde \theta^1 = \theta ~~,~~ \tilde \theta^2 = \phi ~~, \nonumber \\
&
\Ups = a(t) ~~~~~,~~~~~ U^a=0 ~~~~~,~~~~~ \gamma_{ab} \di \tilde{\theta}^a \di \tilde{\theta}^b = a^2(t) r^2 (\di \theta^2 + \sin^2 \theta \, \di \phi^2) ~~,
\eea
where $\eta$ is a conformal time parameter such that $\di \eta= \di t/a$ and $r$ is the radial distance related to $\chi$ by $r = f_K(\chi) \equiv \sin\left(\sqrt{K}\chi\right) / \sqrt{K}$. Here $K$ sets 3 different types of spacelike-hypersurface geometries\,: $K \in \{-1,0,1\} \Rightarrow $ \{open, flat, closed\} Universe, and we authorise the square root to receive negative arguments, given that $\sqrt{-K} = i \sqrt{K}$, and consider the complex expression of the sine function. In the particular case of a flat geometry, one notice that the transformation of the null coordinate is $w = r + \eta$.
The equality $\tau = t$, with $t$ the time of the synchronous gauge, holds at the exact non-perturbative level (see \cite{P2}).

Employing the metric elements of Eq. \rref{FR}, we obtain the expressions of the lensing quantities defined in Sec. \ref{Sec2} within the FLRW geometry for a better understanding of the non-trivial cases. The angular distance is by definition $\bar{d}_A$ and is given after the evaluation of $\det \gamma_{ab} = a^4 r^4 \sin^2 \theta$ and $\det^{ab} \dot{\gamma}_{ab} = 4 a^4 r^2 \left( H(t) r - \frac{1}{a(t)} \right)^2 \sin^2 \theta$ \,:
\beq
\left( \frac{\bar{d}_A}{\bar{u}_{\tau_o}} \right) = a(\tau)^2 r^2 ~~,
\eeq
where we have also used that $r_o = 0$ and $\theta = \theta_o$ in the homogeneous FLRW case.
This result can also be obtained by a direct transformation between GLC and FLRW coordinates. This transformation hence gives $\bar d_A = u_{\tau_o} \, \Gamma(\tau)\left[ w-\int\Gamma^{-1}(\tau) \di \tau \right] / \left(1-\left[\dot\Gamma(\tau)\left( w-\int\Gamma^{-1}(\tau) \di \tau \right)\right]_o \right)$ where $\Gamma = \Gamma(\tau)$ is an arbitrary function that we can identify with the scale factor $a(\tau)$ and $w-\int\Gamma^{-1}(\tau) \di \tau$ corresponds to the conformal radius $r$ from the observer.

By our choice of normalisation the magnification given by Eq. \rref{MagInGLC} is $\mu = 1$. The convergence, vorticity and shear taken from Eq. \rref{LensingCombinationsInGLC} turn out to be trivial\,:  $\left( 1-\kappa \right)^2+\hat\omega^2 = 1$, i.e. $\kappa = 0$ and $\hat{\omega} = 0$, and $\hat\gamma_1^2+\hat\gamma_2^2 = 0$. The optical scalars are given by Eqs. \rref{ThetaHat} and \rref{SigmaHat} and we have\,:
\beq
\hat \theta = \frac{\omega ( -1 + a H r )}{ a^2 r } ~~~~~,~~~~~ | \hat \sigma |^2 = 0 ~~~~~.
\eeq
Finally the use of Eqs. \rref{Phi00} and \rref{Psi0} leads us to\,:
\beq
\Phi_{00} = \frac{\omega^2}{16 a^4 r^2} \left( 16 - \frac{1}{\sin^2 \theta} - 32 a H r + 16 a^2 r^2 (H^2 + \dot{H})  \right) ~~~~~,~~~~~ | \Psi_0 |^2 = 0 ~~~~~.
\eeq

\end{appendices}

\vspace{1cm}

% ----- Bibliography ----- %

%\newpage
%\bibliographystyle{ieeetr}
\bibliographystyle{plain}
\bibliographystyle{utphys}
\bibliography{biblioNov2013}

% ------------------------ %

\end{document}